\DeclareUnicodeCharacter{2212}{-}
\documentclass[journal=jpcafh,manuscript=article]{achemso}
\usepackage{longtable}
\usepackage{multirow}
\usepackage{amsmath}
\usepackage{subfigure}

\usepackage[usenames,dvipsnames]{color}
\usepackage{soul}
\usepackage{bm}
\usepackage{rotating}

\usepackage{amssymb}
\usepackage{titlecaps}
\Addlcwords{a an the of on at with for and in into from by without}
\usepackage[normalem]{ulem} % added to allow striked-out text for editing
\usepackage{xr}
\usepackage{multicol} \usepackage{wrapfig} \usepackage{enumitem}
\usepackage{bm} \usepackage{outlines} %for itemize with subitems
\SectionNumbersOn

\author{Haydar Taylan Turan, Eric Boittier} \affiliation[University of
  Basel]{Department of Chemistry, University of Basel,
  Klingelbergstrasse 80 , CH-4056 Basel, Switzerland.}

\author{Markus Meuwly} \affiliation[University of Basel]{Department of
  Chemistry, University of Basel, Klingelbergstrasse 80 , CH-4056
  Basel, Switzerland.}  \email{m.meuwly@unibas.ch}

\title{Interaction at a Distance: Xenon Migration in Mb}

\begin{document}
% \date{\today}

\begin{abstract}
The transport of ligands, such as NO or O$_2$, through internal
cavities is essential for the function of globular proteins, including
hemoglobin, myoglobin (Mb), neuroglobin, or truncated hemoglobins. For
Mb, internal cavities Xe1 through Xe4 were observed experimentally and
they were linked to ligand storage. The present work determines
barriers for xenon diffusion and relative stabilization energies for
the ligand in the initial and final pocket linking a transition
depending on the occupation state of the remaining pockets from both
biased and unbiased molecular dynamics simulations. It is found that
the energetics of a particular ligand migration pathway may depend on
the direction in which the transition is followed and the occupation
state of the other cavities. Furthermore, the barrier height for a
particular transition can depend in a non-additive fashion on the
occupation of either cavity A or B or simultaneous occupation of both
cavities, A and B. Multiple repeats for the Xe1$\rightarrow$Xe2
transition reveal that the activation barrier is a distribution of
barrier heights rather than one single value which is confirmed by a
distribution of transition times for the same transition. Dynamic
cross correlation maps demonstrate that correlated motions occur
between adjacent residues or through space, residue Phe138 is a gate
for the Xe1$\rightarrow$Xe2 transition, and the cavity volumes
vary along the diffusion pathway, indicating that there is dynamic
communication between the ligand and the protein. The sum of these
molecular-level analyses suggests that Mb is an allosteric protein.
\end{abstract}

\section{Introduction}
Proteins can contain cavities which play potentially functional roles
as they allow transient ligand localization and transport between
neighboring sites. Although cavities have been traditionally regarded
as ``packing defects'' that can destabilize a protein
structure\cite{connolly:1986} they are quite ubiquitous. Directly
mapping out such cavities has been possible for a number of proteins,
including myoglobin (Mb),\cite{tilton:1984} Cytochrome ba3
Oxidase\cite{luna:2012}, T4 lysozyme\cite{liu2008use}, or
interleukin-1$\beta$.\cite{quillin2006determination} For T4 Lysozyme
cavity-creating mutations by replacing various amino acids by alanine
were found to decrease the thermodynamic stability of the protein by
2.7 kcal/mol to 5.0 kcal/mol.\cite{eriksson:1992} For azurin similar
experiments were carried out which also indicated that thermodynamic
stability is decreased upon replacement of wild type residues by
alanine.\cite{gabellieri:2008} Additional analysis revealed that there
is no correlation between thermodynamic stability and flexibility
which may be related to the particular modification sites
chosen. Finally, for hen egg white lysozyme cavity filling mutations
were explored for which the Met12Phe mutant was found to be more
stable than the WT protein by 0.8 kcal/mol.\cite{ohmura:2001} More
recently, cavities were also designed ``de novo'' from computer
modeling and confirmed by X-ray crystallography with the largest such
cavity measuring 520 \AA\/$^3$.\cite{baker:2017}\\

\noindent
In addition to cavities as ligand binding positions within a protein,
such cavities can also be part of a ligand storage and diffusion
network such as for myoglobin or truncated Hemoglobin
(trHb),\cite{Milani:2004,MM.trhb:2012} neuroglobin
(Ngb),\cite{Brunori2007,Kriegl2002,Lutz2009,Nienhaus2010} or
cytoglobin (Cyt).\cite{bolognesi:2004} For myoglobin, four xenon
binding sites (Xe1 to Xe4) in addition to the main binding site
(usually referred to as ``A'' for bound and ``B'' for unbound ligand)
at the heme-iron were characterized experimentally.\cite{tilton:1984}
The physiological relevance of individual sites has been discussed in
the literature and indicates that blocking sites ``B'' and/or ``Xe4''
have profound physiological consequences for the overall rates of
O$_2$ binding to Mb.\cite{olson:2007} Filling the main binding site by
suitable mutations lowers the rates of both ligand recruitment and
dissociation. Furthermore, in a comprehensive mutagenesis study, the
Xe4 cavity was filled by inserting Trp at strategic positions. Such
mutations block access of the ligand to the protein interior. This
leads to increased geminate recombination. However, it was also
reported that the rate of ligand entry is modulated by Trp mutations
in the ``B'' and in Xe4 pockets. Hence, the size of the distal pocket
plays a central role for ligand entry and capture and the two
processes appear to be concerted on the $\mu$s and ms time
scales.\cite{olson:2007} Blocking the Xe1 site was also studied by
either adding Xe gas or by mutation of L89, L104, or F138 to
Trp.\cite{Scott2001} There is little change in the total fraction of
geminate recombination but all these modifications markedly reduce the
amplitude of the slower recombination phase,\cite{Scott2001} possibly
due to rigidification of the iron-out-of-plane motion which gates
ligand rebinding.\cite{bartunik:1999}\\

\noindent
While xenon as a ligand to myoglobin does not have a biological
function it is often used as a convenient probe to map out internal
cavities.\cite{tilton:1984,Nishihara2004,luna:2012} This is also what
xenon is used for in the present work. For proteins with multiple
internal binding pockets the question arises whether and in what way
ligand migration pathways depend on the internal occupation state. In
other words, the question is whether for a transition between cavities
A and B the occupation state of cavity C (empty or filled) affects the
barrier that needs to be overcome. Depending on the answer to this
question, important implications result for protein function and
protein design. As mentioned above, occupation of Xe1 has been found
to markedly reduce the amplitude of the slower recombination rate of a
ligand at the main binding site. Hence, it is expected that a more
comprehensive assessment of the influence of internal occupation state
(by physically placing a ligand - here xenon - or by mutation to
partially or entirely fill an internal site) will provide further
information about the relationship between cavity occupation and
migration pathways and associated energetics.\\

\noindent
The term ``allostery'' - Greek for ``other site'' - used with
reference to molecular-level control of cellular function, was
introduced in 1961 to describe ``interaction at a distance'' involving
two (or multiple) binding sites in a protein.\cite{monod:1961}
However, the concept is older than that, going back to Pauling for
explaining positive cooperativity in binding of molecular oxygen to
Hemoglobin.\cite{pauling:1935} This model eventually developed into
the ``Koshland-Nemethy-Filmer'' (KNF)\cite{koshland:1966} as opposed
to the model by ``Monod, Wyman and Changeux'' (MWC) which provided an
alternative view of allostery. When applied to oxygen binding to
hemoglobin, the MWC involves only changes at the quaternary structural
level whereas KNF is only concerned with tertiary structural
changes. Allostery is also a fundamental regulatory mechanism that
plays a role in various biological processes such as gene
regulation\cite{tsai2008allostery}, signal
transduction\cite{changeux2005allosteric}, and
metabolism\cite{link2014advancing}. Ligand binding is one of the
important ``perturbations" that causes allosteric regulation.
Previous models of allostery, such as KNF and MWC, were based on
crystallographic observations and were limited in focus to
perturbations at the tertiary and quaternary structural
level.\cite{cui:2008,motlagh:2014,guo:2016} Due to advancements in NMR
spectroscopy,\cite{strotz:2020} alongside developments in molecular
dynamics (MD)
simulation,\cite{sethi:2009,Chennubhotla:2006,atilgan:2004} insights
into allosteric regulation induced by ligand binding at the molecular
level have emerged.  These approaches have benefited from enhanced
sampling methods in MD, as well as graph-based community and network
detection algorithms which have been shown to be effective at
identifying important residues and transition pathways to further the
understanding of allostery at a molecular
level.\cite{Chennubhotla:2006, hayatshahi:2019}\\

\noindent
The present work is structured as follows. First, the computational
methods including the MD simulations and their analysis is
presented. This is followed by a discussion of the migration
energetics for Xe depending on the occupation state of the remaining
cavities. Next, the correlated motions within Mb are characterized and
volume changes of the pockets are analyzed. Finally, the results are
discussed in a broader context of allostery and conclusions are
drawn.\\

\section{Computational Methods}

\subsection{Molecular Dynamics}
All molecular dynamics (MD) simulations were performed using the
CHARMM\cite{Brooks1983} software with CHARMM22 force field, with
$R_{\rm min}$ = 2.05 \AA\/ and $\epsilon_{ij} = -0.42$ kcal/mol for
the Xe atom. This radius is somewhat smaller than the one used
previously ($R_{\rm min}$ = 2.24
\AA\/\cite{cohen2006imaging}). Myoglobin (Mb) was placed in the center
of the solvent box with dimensions $65.2 \times 59.0 \times 46.6$
\AA\/$^3$, and maintained there with a 1 kcal/mol center of mass
constrain. MD simulations were started following 500 steps of steepest
descent and 500 steps of Adopted Basis Newton−Raphson minimization
then the system was heated to 300 K and equilibrated for 50 ps in
$NVT$ ensemble. Then, 1 ns of $NVE$ was carried out with leapfrog
integrator\cite{hairer2003geometric} using a time step of $\Delta t =
1$ fs, and all bonds involving hydrogen atoms were constrained using
SHAKE.\cite{ryckaert1977numerical} Non-bonded interactions were
treated with a switch function\cite{steinbach1994new} between 12 and
16 \AA\/. Xe atoms not involved in transitioning between neighboring
pockets are retained in their given pockets by a harmonic constrain
with a force constant of 10 kcal/mol. \\

\begin{figure}[H]
\begin{center}
\includegraphics[width=0.70\linewidth]{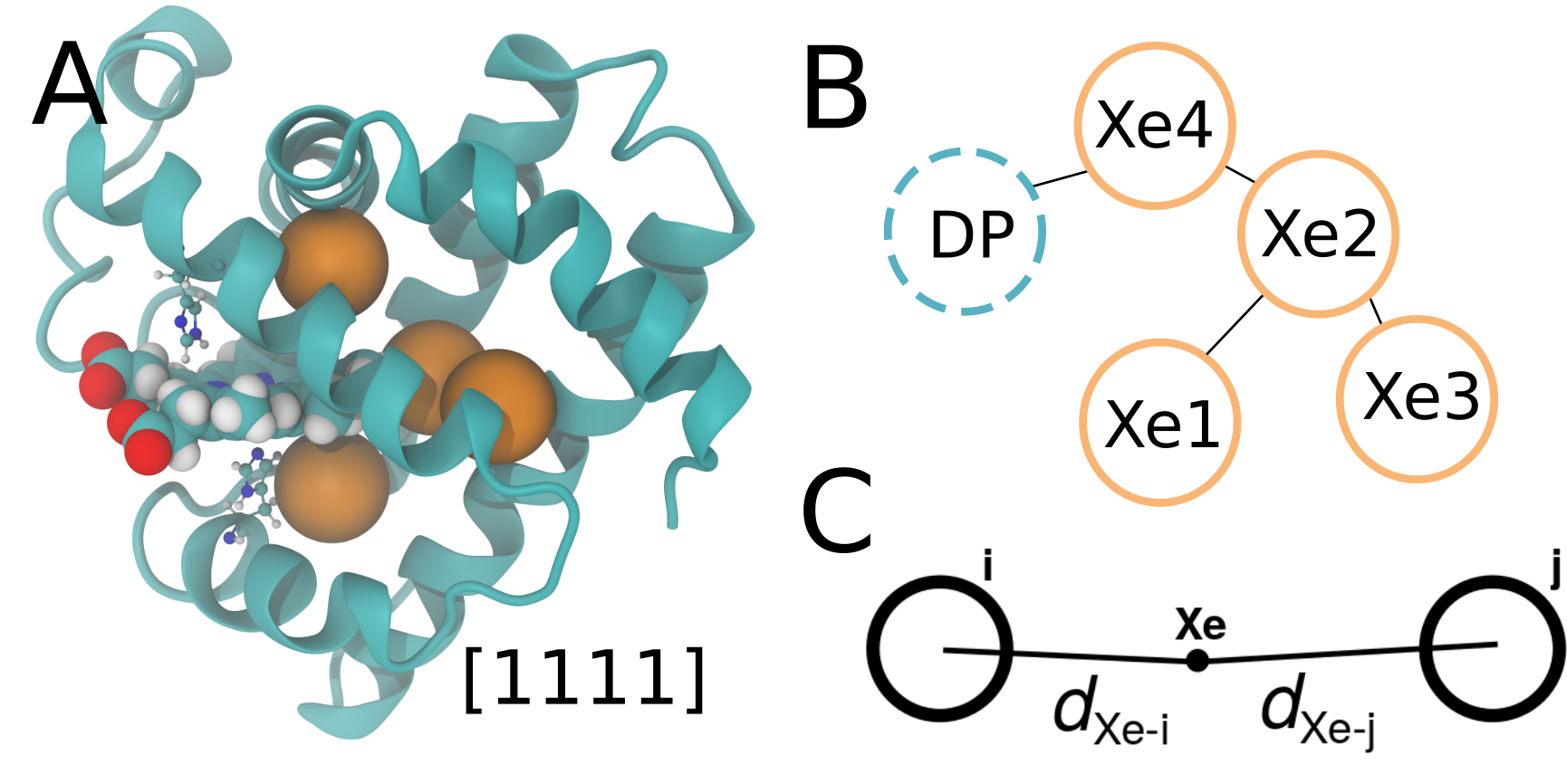}
\caption{Panel A. The structure of myoglobin with four internal
  pockets (Xe1 - Xe4) occupied with xenon atoms (orange
  spheres). Panel B. An idealized graph showing possible migration
  pathways between pockets, including the distal pocket (DP), which
  are investigated using umbrella sampling. Panel C. Definition of the
  reaction coordinates $r^{\rm A}_{\rm c}$ and $r^{\rm B}_{\rm c}$
  based on $d_{\rm Xe - i}$ and $d_{\rm Xe - j}$ distances.}
\label{fig1}
\end{center}
\end{figure}

\noindent
All simulations were initialized from an initially equilibrated fully
occupied protein (1111) with pockets Xe1 to Xe4 occupied by one xenon
atom each. A binary system is used in the following to identify the Xe
occupation state of Mb. For example, 0100 represents Mb with Xe
present in Xe2, with pockets Xe1, Xe3, and Xe4 empty, whereas 1011
represents a system in which Xe1, Xe3, and Xe4 host a Xe atom.  The Xe
atoms were constrained to their given pockets and the initial 1111
state was equilibrated for 2 ns in the $NVT$ ensemble.  From this
structure, 16 different Xe occupation systems were generated, covering
all possible occupation states with regards to cavities Xe1, Xe2, Xe3,
and Xe4 \cite{tilton:1984}, see Figure \ref{fig1}. Each system was
again heated and equilibrated for 2 ns in the $NVT$ ensemble with the
Xe atoms constrained to their pockets. For Xe in DP, the last
structure from the umbrella simulations in given state is extracted,
and equilibrated for 2 ns in the $NVT$ ensemble with the Xe atom
constrained in DP. The definitions for the cavities were those from
prior characterizations of the Xe pockets as shown in Table
\ref{tab:pockets}.\cite{Tilton1986,Bossa2005,Park2006,Anselmi2008,MM.trhb:2012}
The Xe pockets and HEME residue in myoglobin is shown in
Figure~\ref{fig1}. The connection between pockets are labelled with
black lines.\\

\begin{table}[h]
\begin{center}
\begin{tabular}{  c ||  c  }
 \hline
 Cavity & Molecular definition\\
 \hline
 Xe1 & Leu89, His93, Leu104, Phe138, Ile142, Tyr146 and HEME\\
 Xe2 & Leu72, Ile107, Ser108, Leu135, Phe138, Arg139 and HEME\\
 Xe3 & Trp7, Thr75, Leu76, Gly80, His82, Ala134, Leu137, Phe138\\
 Xe4 & Gly25, Ile28, Leu29, Gly65, Val68, Leu69, Leu72, Ile107, Ile111\\
 Distal pocket (DP)& Leu29, Phe46, Hse64, Val68, Ile107\\
 \hline
\end{tabular}
\end{center}
\caption{Name of each cavity within Mb along with molecules and
  residues which line the pocket as defined by previous
  research.\cite{frauenfelder:2001,Anselmi2011}}
\label{tab:pockets}
\end{table}

\subsection{Umbrella Sampling Simulations}
Direct sampling of Xe atoms between neighboring pockets may be slow
due to the appreciable energy barriers between the pockets ($> 5$
kcal/mol). Therefore, umbrella sampling (US) simulations were also
used to sample the reaction pathway between the pockets. The reaction
coordinate chosen include 1) the distance between the Xe atom and the
center of mass of the initial pocket $i$ ($d_{\rm Xe - i}$)
i.e. $r^{\rm A}_{\rm c}$ = $d_{\rm Xe - i}$ and 2) the difference of
distance between the Xe and center of mass of initial pocket $i$
($d_{\rm Xe - i}$), and distance between between the Xe and center of
mass of final pocket $j$ ($d_{\rm Xe - j}$), hence $r^{\rm B}_{\rm c}$
= $d_{\rm Xe - i} - d_{\rm Xe - j} $. Simulations were carried out for
equidistant windows with $\Delta r_{\rm}$ = 0.1 \AA\/ and $k_{\rm
  umb}$ = 25 kcal/mol.  \\
  
\noindent
For discussing all transitions on the same footing, the $r_{\rm c}$
values were scaled to map $r^{\rm A}_{\rm c}$ and $r^{\rm B}_{\rm c}$
to the interval $[0.0,1.0]$, as shown in Eq.~\ref{eq:scale}, 
\begin{equation}
    r_{\rm c} (x) = \frac{r^{\rm A}_{\rm c} (x) - {\rm min}(r^{\rm
        A}_{\rm c})}{\rm{max}(r^{\rm A}_{\rm c}) - {\rm min}(r^{\rm
        A}_{\rm c})},
\label{eq:scale}
\end{equation}
and visualized in Figure~\ref{fig1}C. In this way, the reaction
coordinate begins inside the minimum of the starting pocket and
extends to the minimum of the terminal pocket.\\

\noindent
US simulations were carried out in a sequential manner for both of
$r^{\rm A}_{\rm c}$ and $r^{\rm B}_{\rm c}$, each window was simulated
for 250 ps, and windows statistics were accumulated after
equilibration for 20 ps. Statistics were merged from all windows to
yield a one-dimensional potential of mean force (PMF) by using the
weighted histogram analysis method
(WHAM)\cite{Kumar1992,souaille.comphycom.2001.us} with a tolerance of
0.001. \\

\subsection{Dynamical Cross-Correlation Maps}
The dynamical cross-correlation maps (DCCM) and difference dynamical
cross-correlation maps \cite{ichiye1991collective,arnold1997molecular}
($\Delta$DCCM) were calculated to quantitatively characterize the
effects of Xe transition inside the protein on the dynamics using the
Bio3D package.\cite{grant2006bio3d} Dynamic cross-correlation maps
matrices and coefficients

\begin{equation}
    \label{dccm_func}
    C_{ij} = \langle\Delta r_{i} \cdot \Delta r_{j}\rangle /
    (\langle\Delta r_{i}^{2} \langle\Delta r_{j}^{2}\rangle)^{1/2}
\end{equation} 
were determined from the position of C$\alpha$ in amino acids $i$ and
$j$ with positions $r_{i}$ and $r_{j}$. $\Delta r_{i}$ and $\Delta
r_{j}$ determine the displacement of the $i$th C$\alpha$ from its
average position throughout the trajectory. One should note that DCCM
characterizes the correlated ($C_{ij}$ $>$ 0) and anti-correlated
($C_{ij}$ $<$ 0) motions in a protein whereas $\Delta$DCCM provide
information about which couplings are affected by the presence of Xe
atoms in a given state with respect to unoccupied protein.\\

\subsection{Volume Analysis}
The SURFNET package\cite{Laskowski1995} was used for detecting Mb
cavities Xe1, Xe2, Xe3, and Xe4 and calculating their volumes. Two
different probe size $r_{\mathrm{min}} = 1.2$ \AA\/ and
$r_{\mathrm{max}} = 3.5$ \AA\/ were used to determine the volume of
the cavities.\cite{Bossa2005,Anselmi2011}. The distribution of
volumes, $p(V)$, and their median was determined for each US window.\\

\section{Results}
First, the potentials of mean force are discussed to provide an
overview of the influence of pocket occupation on the ligand
dynamics. Next, the results from unbiased simulations are presented
and discussed in the context of the US simulations. Then, the global
dynamics of Mb in different xenon occupation states is discussed by
means of dynamical cross correlation maps. This is followed by a
broader discussion linking the energetics for ligand migration with
structural changes and eventually allostery in Hb.\\

\subsection{Potentials of Mean Force Starting from Mb(1111)}
Initially, all umbrella sampling simulations were started from fully
occupied Mb (1111) by removing surplus Xe atoms depending on the state
that was investigated. Next, all remaining Xe atoms that were not part
of the transition for which US was carried out were weakly constrained
to the center of mass of their respective pockets. Such a preparation
will initially not be structurally adapted to the respective initial
occupation state, i.e. additional structural relaxation is
possible. First, the profiles following this preparation are discussed
and in a next step the consequences of additional relaxation are
considered.\\

\begin{figure}[H]
\begin{center}
\includegraphics[width=\linewidth]{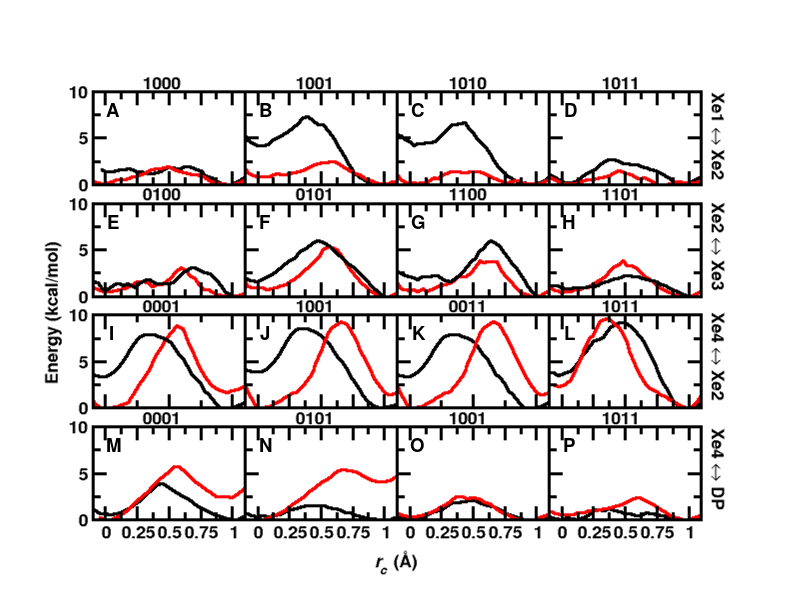}
\caption{The PMFs for the Xe1$\leftrightarrow$Xe2,
  Xe2$\leftrightarrow$Xe3, Xe4$\leftrightarrow$Xe2,
  Xe4$\leftrightarrow$DP transition in various systems. The forward
  transitions are in black and the reverse transitions are in red.}
\label{fig2}
\end{center}
\end{figure}

\noindent
All potentials of mean force for the different occupation states are
summarized in Figure \ref{fig2}. From top to bottom the
Xe1$\leftrightarrow$Xe2, Xe2$\leftrightarrow$Xe3,
Xe4$\leftrightarrow$Xe2, and Xe4$\leftrightarrow$DP transitions are
considered for all possible occupation states for the remaining
pockets. Before discussing individual transitions in some more detail
a broader overview is given. It is noted that the free energy barriers
for the forward and reverse direction do not need to be identical. For
example, for the Xe1$\leftrightarrow$Xe2, with (1011) occupation
(Figure \ref{fig2}D) the initial and final state are energetically
equivalent but the barrier heights for the forward and the reverse
process differ. Alternatively, there are transitions such as
Xe4$\leftrightarrow$DP with (1001) occupation for which the energy
profiles for forward and reverse transition overlap. Also, it is noted
that for certain profiles the differential stabilization of the
reactant and product states can interconvert depending on the
direction in which the scan is carried out, e.g. panels I and J in
Figure \ref{fig2}.\\

\noindent
{\it The Xe1$\leftrightarrow$Xe2 transition:} For the 1000 and 1011
occupation states the stabilization energy with Xe in Xe1 and Xe2 is
similar. The forward and reverse barriers, $\Delta G_{\rm f}$ and
$\Delta G_{\rm r}$, are comparable for 1000 but differ by $\sim 0.9$
kcal/mol for 1011. With 1011 occupation the activation barrier is
higher ($\Delta G_{\rm f} = 2.44$ kcal/mol) and earlier ($r_{\rm c} =
0.35$) for the forward transition compared with the reverse transition
($\Delta G_{\rm r} = 1.35$
kcal/mol and $r_{\rm c} = 0.35$) whereby the initial and final states
are isoenergetic, see Figure \ref{fig2}D. For the 1001 and 1010
occupation the energy for the state with Xe in Xe1 is higher when
running the US simulation in the forward direction compared with
simulations in the reverse direction.\\

\begin{figure}[h!]
\begin{center}
  \includegraphics[width=0.8\linewidth]{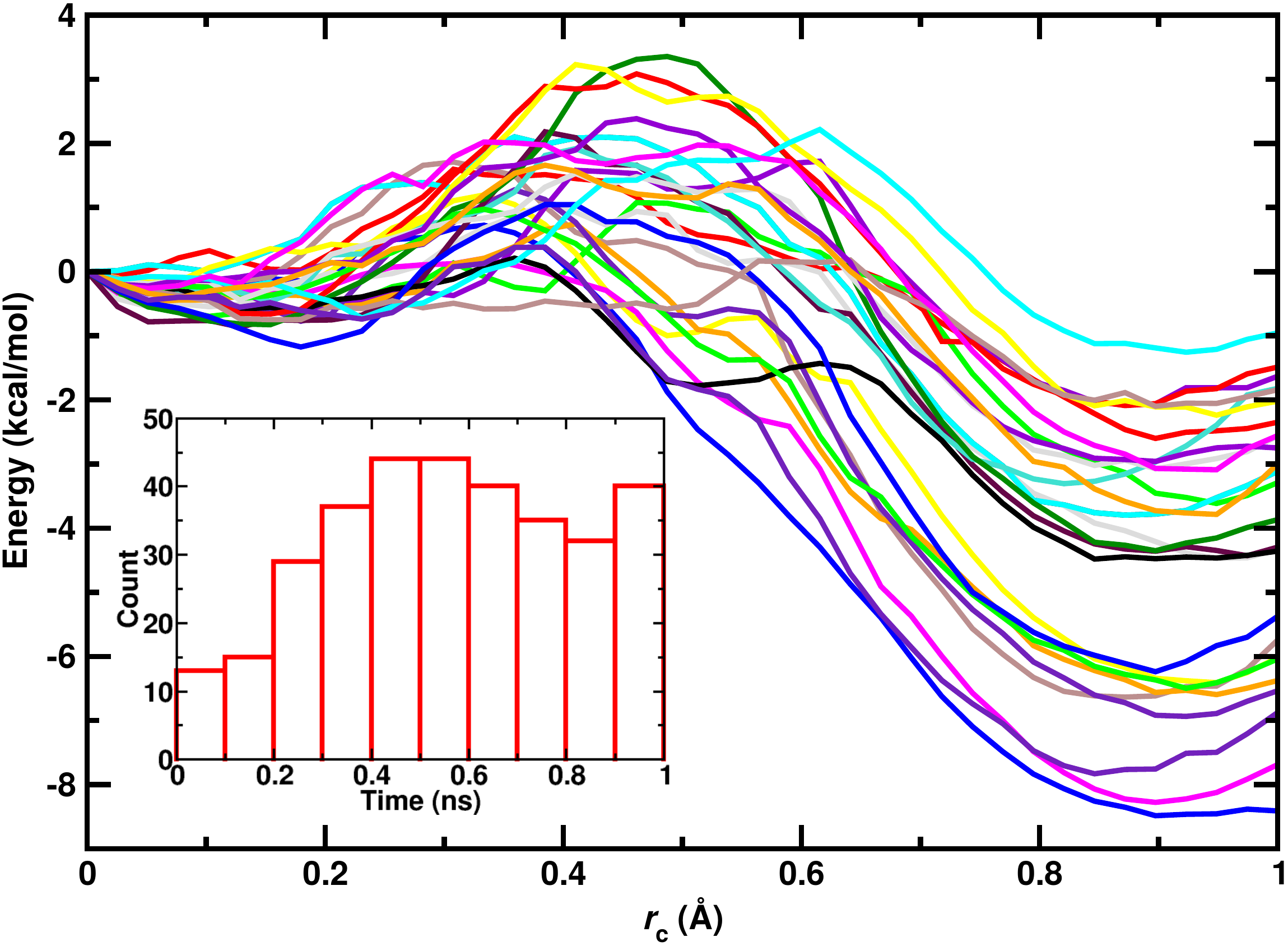}
\caption{The PMF of the Xe1$\rightarrow$Xe2 transition in 1001. The
  original PMF (black), as shown in Figure~\ref{fig2}, is compared to
  the PMF for the 27 repeat simulations along the same $r_{\rm c}$
  (red). Inset) The distribution of transition times for the
  Xe2$\rightarrow$Xe1 transition in 0101 with bin size of 100 ps for
  the transition times. Overall, 2500 independent trajectories 1 ns in
  length were run, leading to 329 transitions.}
\label{fig3}
\end{center}
\end{figure}

\noindent
To assess how representative the single PMFs in Figure \ref{fig2} are,
a larger number of independent US simulations was carried out for one
specific transition in one occupation state. Repeating USs from
different initial conditions for the Xe1$\rightarrow$Xe2 transition
with 1001 occupation are shown in Figure \ref{fig3} and yield barrier
heights ranging from 0.56 kcal/mol to 4.2 kcal/mol. From close to 30
repeats of the same US simulation the averaged forward barrier height
was $2.04 \pm 0.86$ kcal/mol. Hence, when starting from the same
initial structure but with independent initial preparation by starting
the heating from a different random seed for the velocities, the
barrier height (here $\Delta G_{\rm f}$) can vary appreciably. On the
other hand, all profiles agree in that the structure with Xe in Xe1 is
higher in energy than that with Xe in Xe2. Thus, structural
heterogeneity is reflected in a distribution of Xe migration barrier
heights and similar findings are expected for all other free energy
profiles shown in Figure \ref{fig2}.\\

\noindent
{\it The Xe2$\leftrightarrow$Xe3 transition:} For this transition and
all occupation states the energy levels for the initial (Xe in Xe2)
and final (Xe in Xe3) states are independent on the direction in which
the transition was probed. Typically, the state with Xe in Xe2
($r_{\rm c} = 0.0$) is slightly higher in energy than for Xe in Xe3
but within the estimated accuracy of the simulations of $\pm 1$
kcal/mol (see e.g. Figure \ref{fig3}), the two states can be
considered isoenergetic. For the 0100 occupation, the PMF is
essentially flat between Xe in Xe2 and the TS with $\Delta G_{\rm f} =
1.6$ kcal/mol and $\Delta G_{\rm r} = 1.8$ kcal/mol and a TS at
$r_{\rm c} = 0.30$ and $r_{\rm c} = 0.35$ in the two directions,
respectively. For 0101 the reverse barrier differs insignificantly by
$\Delta \Delta G_{\rm r} = 0.03$ kcal/mol for the forward and the
reverse directions and the TS occurs at slightly different reaction
coordinate ($r_{\rm c} = 0.48$ and $r_{\rm c} = 0.6$,
respectively). More importantly, however, the barrier heights depend
on whether pocket Xe4 is empty (0100) or occupied (0101). The
difference in the free energy barrier is about a factor of
two. Similarly, for 1100 the barrier heights are larger than for 0100
whereas for 1101 the barriers are more reminiscent of those of 0100
occupation. This suggests not only that the occupation states of the
remaining pockets can affect barrier heights for ligand migration
between two given pockets but also that the effect is not necessarily
additive.\\

\noindent
{\it The Xe4$\leftrightarrow$Xe2 transition:} Barriers for this
transition are uniformly high. Except for the 1011 occupation state
further relaxation and adaptation may be possible for scans in both
directions. This is evidenced by the fact that for the
Xe4$\rightarrow$Xe2 direction with Xe in Xe2 ($r_{\rm c} = 1$) is more
stable for 0001, 1001, and 0011, whereas the reverse is true for
moving Xe in the opposite direction Xe4$\leftarrow$Xe2.\\

\noindent
{\it The Xe4$\leftrightarrow$DP transition:} The occupation of
neighboring pockets significantly lowered the $\Delta G_{\rm f}$ for
Xe4$\leftrightarrow$DP (bottom line in Figure~\ref{fig2}) in 0101
(Fig.~\ref{fig2}N), 1001 (Fig.~\ref{fig2}O), and 1011
(Fig.~\ref{fig2}P). Other than for the forward transition, $\Delta
G_{\rm r}$ decreased for 1001 and 1011 compared with 0001. The
activation barrier decreased from 4 kcal/mol in 0001
(Fig.~\ref{fig2}M) to 1.1 kcal/mol in 1011 ($\Delta G_{\rm r} = 5.7$
kcal/mol). However, occupation of only the Xe2 pocket (0101)
destabilized the Xe4 pocket by $\sim 1$ kcal/mol and had no effect on
$\Delta G_{\rm r}$. The activation barriers for reverse transitions
feature late transition states ($r_c = 0.65$) with the barrier closer
to DP for 0001, 0101, and 1011 states.\\

\noindent
Overall, the Xe4$\leftrightarrow$Xe2 transition is less likely to
occur compared to all other transitions. The longer distance between
these two pockets decreases the probability for this
transition. Occupation of internal pockets can modulate barrier
heights (0100 vs. 0101 for Xe2$\leftrightarrow$Xe3) and the effect of
pocket occupation does not need to be additive (compare $\Delta G$ for
0101, 1100, and 1101 for Xe2$\leftrightarrow$Xe3).\\

\subsection{Further Relaxation for the Xe4$\leftrightarrow$DP Transition}
\begin{figure}[H]
\begin{center}
  \includegraphics[width=\linewidth]{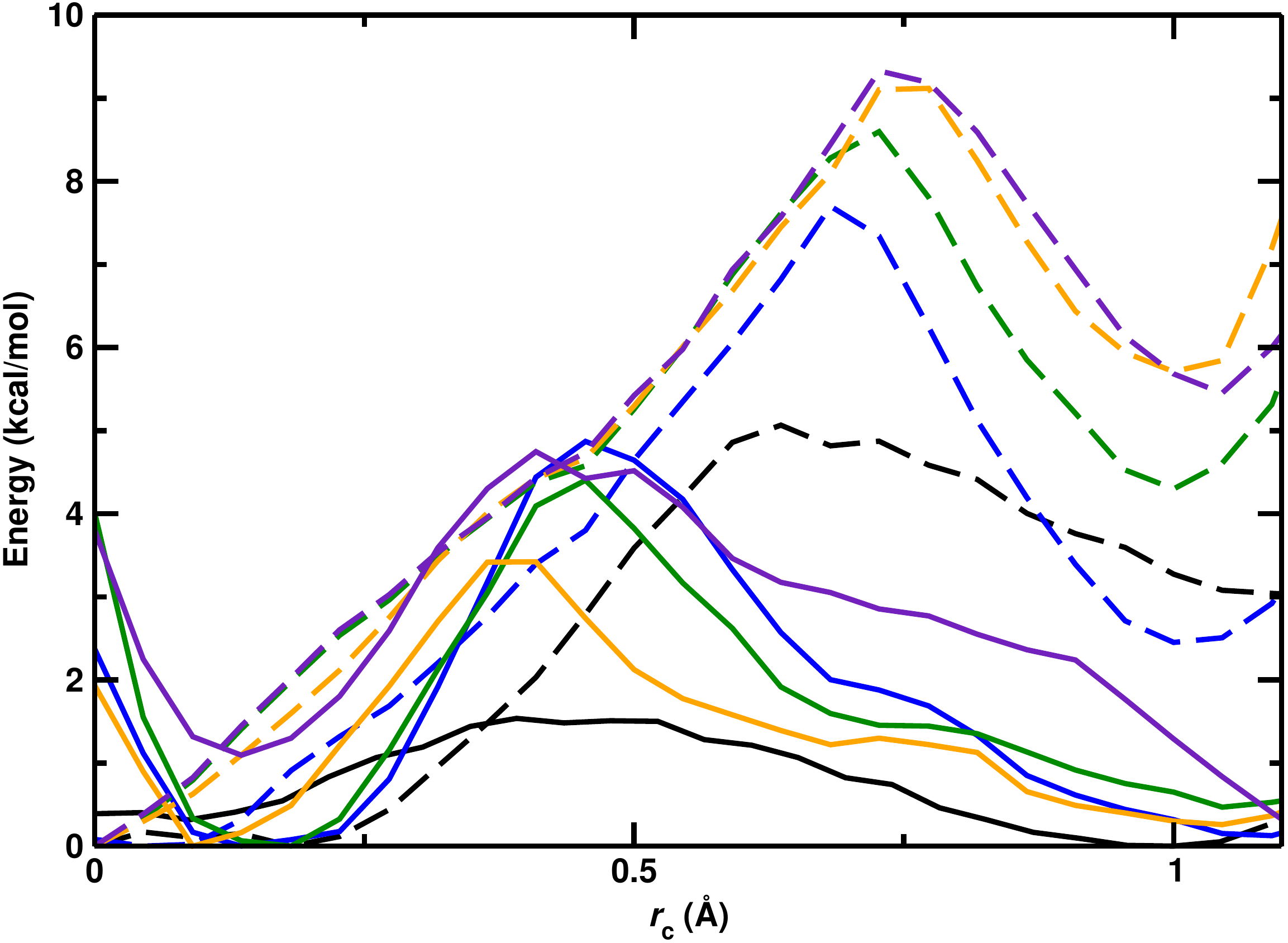}
\caption{The PMF of the Xe4$\leftrightarrow$DP transition for
  0101. PMFs in the forward direction (Xe4$\rightarrow$DP) are the
  solid lines whereas the PMFs in the reverse direction
  (Xe4$\leftarrow$DP) are shown as dashed lines.  The final structure
  of each scan is the starting point for the following scan in the
  opposite direction. The scan order is: black - blue - green - orange
  - indigo. The state with Xe in Xe4 ($r_{\rm c} = 0$) is the zero of
  energy for all scans.}
\label{fig4}
\end{center}
\end{figure}

For 0001 and 0101, the free energy of the states with Xe in pocket DP
differs whether the transition is followed in the forward (black) or
the reverse (red) direction, see Figure \ref{fig2}M and N. As was
already seen for multiple independent US simulations for the
Xe1$\rightarrow$Xe2 transition, variations in barrier heights and
relative stabilization can be expected for repeated scans of the same
transition. To probe the relevance of additional relaxation along the
reaction coordinate, Xe migration in 0101 was repeatedly followed in
the forward (Xe4$\rightarrow$DP) and reverse direction
(DP$\rightarrow$Xe4). For this, the final structure of each scan was
the starting geometry for the following scan in the opposite
direction. Overall, 5 scans in the forward and in the reverse
direction were carried out. The results are shown in Figure~\ref{fig4}
with the forward PMFs as solid lines and those for the reverse scans
as dashed lines in order black - blue - green - orange - indigo.\\

\noindent
In the forward direction, the differential stabilization for Xe in
both states was close to zero. However, the activation barriers
increased from 1.5 kcal/mol to 4.87, 4.40, 3.42, and 4.74 kcal/mol for
blue, green, orange, and indigo scans, respectively. Conversely, for
the reverse scan Xe in DP ($r_{\rm c} = 1$) is always higher in energy
compared with Xe in Xe4 ($r_{\rm c} = 0$). The differential
stabilization changed from 3.0 to 2.5, 4.3, 5.8, and 5.5 kcal/mol for
black, blue, green, orange, and indigo scans, respectively. In
addition, the activation barriers changed from 1.30 to 5.24, 4.30,
3.41, and 3.87 kcal/mol for black, blue, green, orange, and indigo
scans, respectively. These results show the importance of the initial
configuration (see also figure \ref{fig3}) as the structural
(re)arrangements of nearby residues can promote or impede Xe
migration. Furthermore, it is also demonstrated that the differential
stabilization and activation barrier in the forward and reverse
direction for given occupation state (here 0101) can depend on the
direction in which the transition is probed.\\

\noindent
The activation barriers found in the present work are largely
consistent with previous studies. The DP$\rightarrow$Xe4 for the 0001
transition has a barrier of 5.7 kcal/mol compared with 5.4
kcal/mol\cite{ceccarelli2008co} calculated for CO diffusion. Based on
implicit sampling\cite{Cohen2006} barrier heights for Xe diffusion
along different internal paths were determined. For the empty protein
the barrier between DP and Xe4 is between 5 and 6 kcal/mol with the
state with Xe in Xe4 lower in energy. Similarly, for
Xe2$\rightarrow$Xe1 in 0100, the calculated value of 1.5 kcal/mol is
consistent with 1.4 kcal/mol calculated from meta-dynamics simulations
using CO as the ligand.\cite{nishihara2008search}. Additional barriers
were also determined from this approach but it was stated that the
barrier heights were overestimated as a consequence of the implicit
ligand sampling approach. On the other hand, the activation barrier of
4.8 kcal/mol found here for the Xe4$\rightarrow$Xe2 transition in 0001
were somewhat higher compared to the previous studies
2.9\cite{ceccarelli2008co} and 2.7
kcal/mol.\cite{nishihara2008search}. When comparing with earlier
efforts one should note that the barrier heights can depend on the
definition of the reaction coordinate, and the initial
structure/preparation of the system used for the simulations.\\

\subsection{Unbiased Simulations}
Multiple unbiased simulations were performed to compare with the free
energy barriers from the US simulations, see Figure \ref{fig2}. For
this, the transition times $\tau$ for the Xe2$\rightarrow$Xe1
transition for 0101 occupation were determined from 2500 independent
trajectories, each 1 ns in length. Out of those, in 329 runs the
transition of interest occurred which corresponds to 13 \% of the
trajectories. The transition time is determined as the time difference
between the start of the simulation and the first time Xe was in the
Xe1 pocket. Xe is considered to reside in the pocket when the distance
between Xe and the center of the pocket is lower than 4 \AA\/. The 4
\AA\/ threshold chosen due to the fact that 4 \AA\/ sphere would have
volume of $\sim 256$ \AA\/$^3$ which is the volume of the empty Xe1
pocket.\\

\noindent
The distribution of transition times is reported in the inset of
Figure \ref{fig3}. The average transition time is $\sim 500$ ps which
corresponds to a free energy difference of $\sim 3.5$ kcal/mol based
on transition state theory, consistent with 2.5 kcal/mol from the
umbrella sampling simulations, see Figure \ref{fig2}B. However, the
actual transition times cover a range from 68 ps to 1 ns and the fact
that there is a distribution of transition times is also reflected in
the distribution of barrier heights that was found for the repeat US
simulations in Figure \ref{fig3}.\\

\noindent
The results in inset of Figure~\ref{fig3} show that there is not one
specific transition time and that the values for $\tau$ do not follow
a ``simple'' distribution, e.g. a Gaussian, because the underlying
process is one that is rather characterized by ``waiting times'', at
least for short $\tau$. This reflects the fact that depending on the
initial configuration from which the simulation was started more or
less structural arrangements and fluctuations are required to promote
diffusion of Xe from the initial to the final pocket.\\

\subsection{Correlated Motions from Dynamic Cross Correlation Maps}
For a more atomistically refined picture regarding the underlying
motions in the proteins, DCCMs provide a basis to determine the
importance of transition direction and neighboring pocket occupation
on the overall and site-specific dynamics. As the PMFs in
Figure~\ref{fig2} show, the profile along the reaction coordinate
between the initial and final state does not need to be
symmetrical. Hence, the (local) protein dynamics accompanying a
transition may differ in the forward and the reverse direction. This
is indeed what is observed in the DCCM maps. Such maps were
constructed for specific transitions and occupations from US and from
unbiased simulations and are discussed in the following.\\

\begin{figure}[H]
\begin{center}
\includegraphics[width=0.4\linewidth]{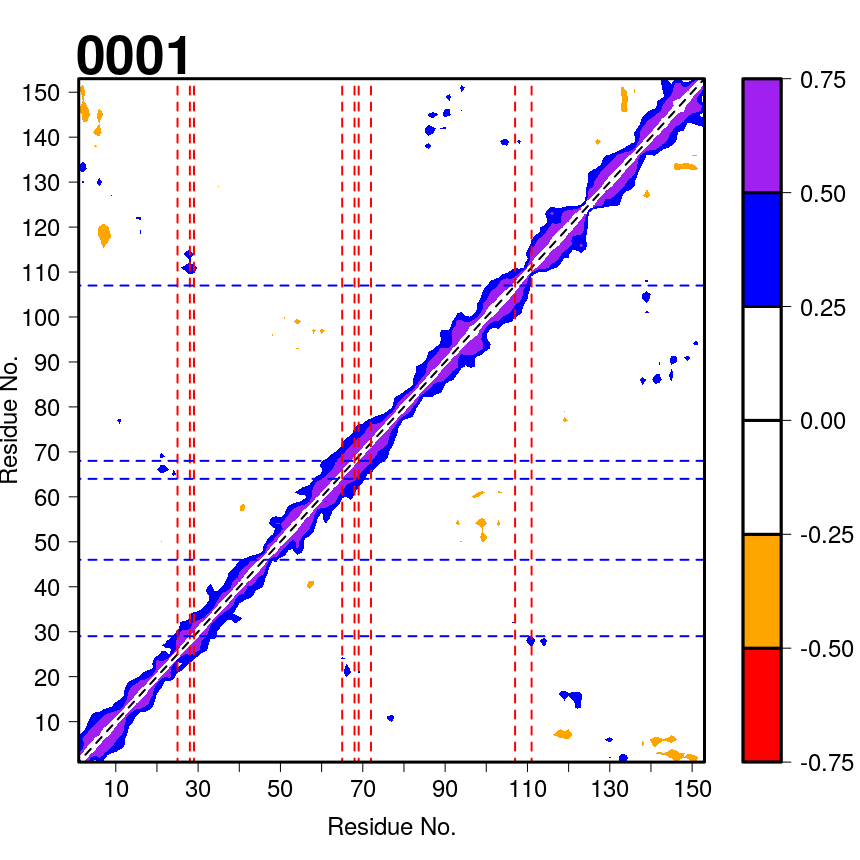}
\includegraphics[width=0.4\linewidth]{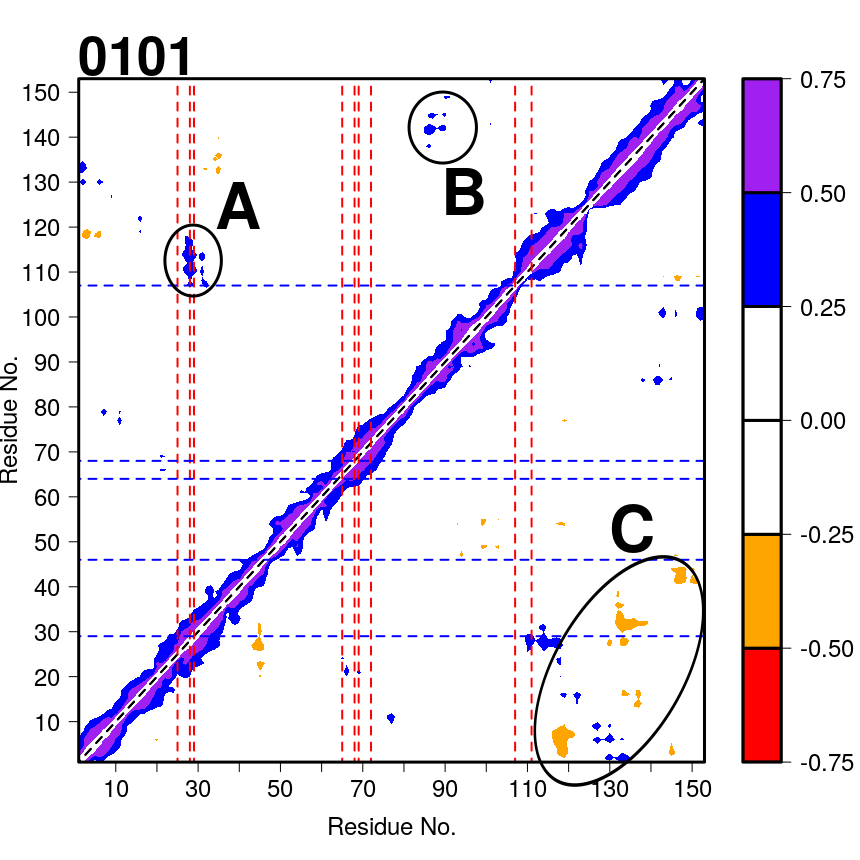}
\includegraphics[width=0.4\linewidth]{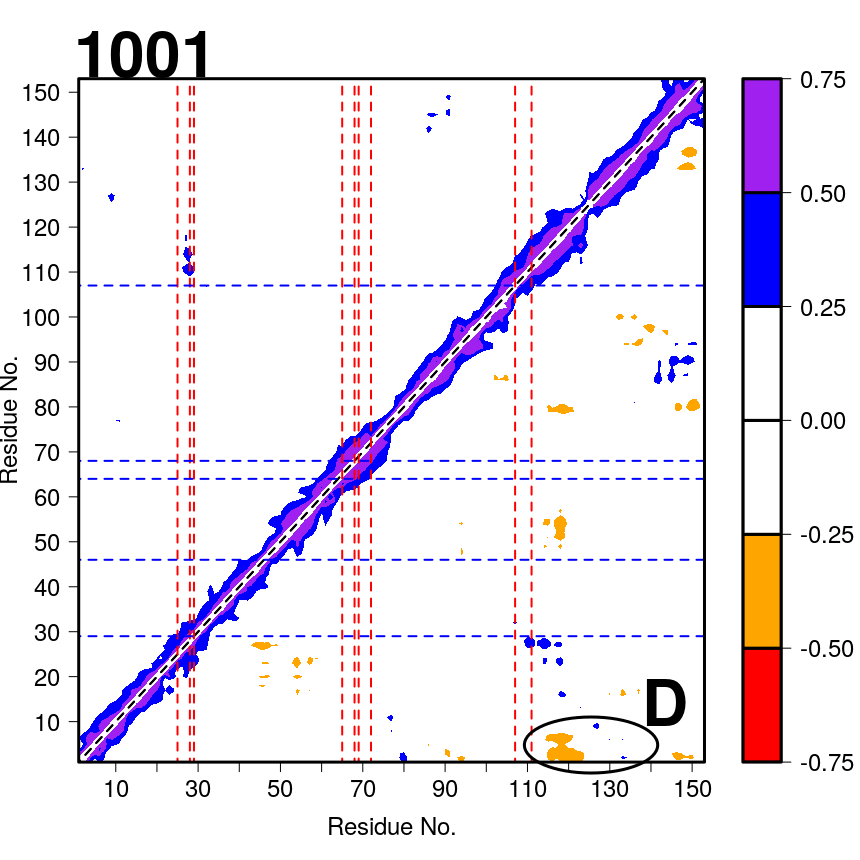}
\includegraphics[width=0.4\linewidth]{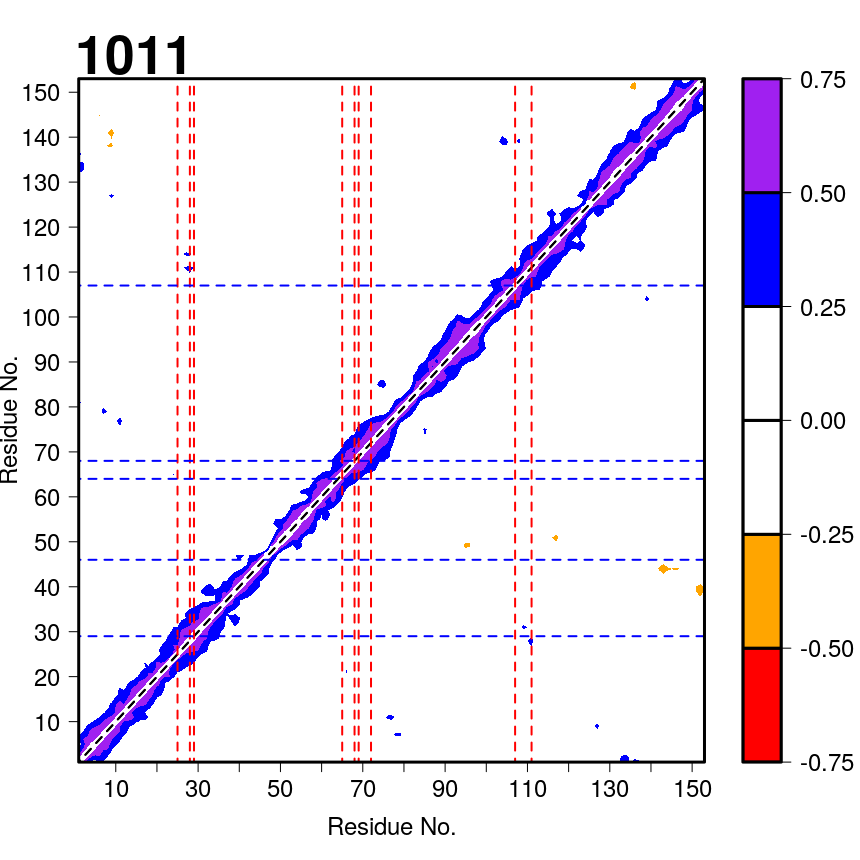}
\caption{The DCCM of Xe4$\rightarrow$DP (upper left triangle) and
  Xe4$\leftarrow$DP (lower right triangle) transition in 0001, 0101,
  1001, 1011. Vertical red lines indicate the residues that form the
  Xe4 pocket whereas horizontal blue lines indicate the residues that
  form the DP. Particular features in the maps are labeled from A to D
  with circles around them }
\label{fig5}
\end{center}
\end{figure}

\noindent
For the Xe4$\leftrightarrow$DP transition (for PMFs see Figure
\ref{fig2}M to P and Figure \ref{fig4}) the motions in the reverse
direction (Xe4$\leftarrow$DP; lower right triangles in Figure
\ref{fig5}) compared with the forward direction (Xe4$\rightarrow$DP;
upper left triangles in Figure \ref{fig5}) were more
correlated/anti-correlated.  Correlated motions can occur between
residues adjacent to one another or through space, in particular if
the motion involves entire secondary structural motifs.\\

\noindent
For the Xe4$\leftrightarrow$DP transition, feature A in
Figure~\ref{fig5} shows that correlated motions between the Gly25,
Ile28, and Leu29 (part of pocket Xe4) with Ile107 to Ser117 are
present for both forward and reverse directions. Ile107 is part of
both, pockets Xe2 and Xe4. These residues are spatially close to each
other and underscore coupling of Xe movement to the local protein
dynamics. Correlated motions are also affected by the Xe occupation of
neighboring pockets. Feature B in Figure~\ref{fig5}, between Leu135
and Lys146 and Pro88 to Ser92 is present in 0001, 0101, and 1001 with
different intensities, but not for 1011. For 1001 there are multiple
correlated and anti-correlated motions for residues between
[N-terminus,Lys50] and [Ala110,C-terminus] (Feature C). States with
higher activation barrier (0001 and 0101, see Figure \ref{fig2}),
showed more intense peaks in the DCCM maps. Anti-correlated motions
between Val1 to Gln8 and Leu115 to Phe123 (feature D) shows spatially
long-range motion correlation, which means that the effect of local
dynamics can affect the global dynamics depending on the particular
occupation state.\\

\begin{figure}[h!]
\begin{center}
\includegraphics[width=0.65\linewidth]{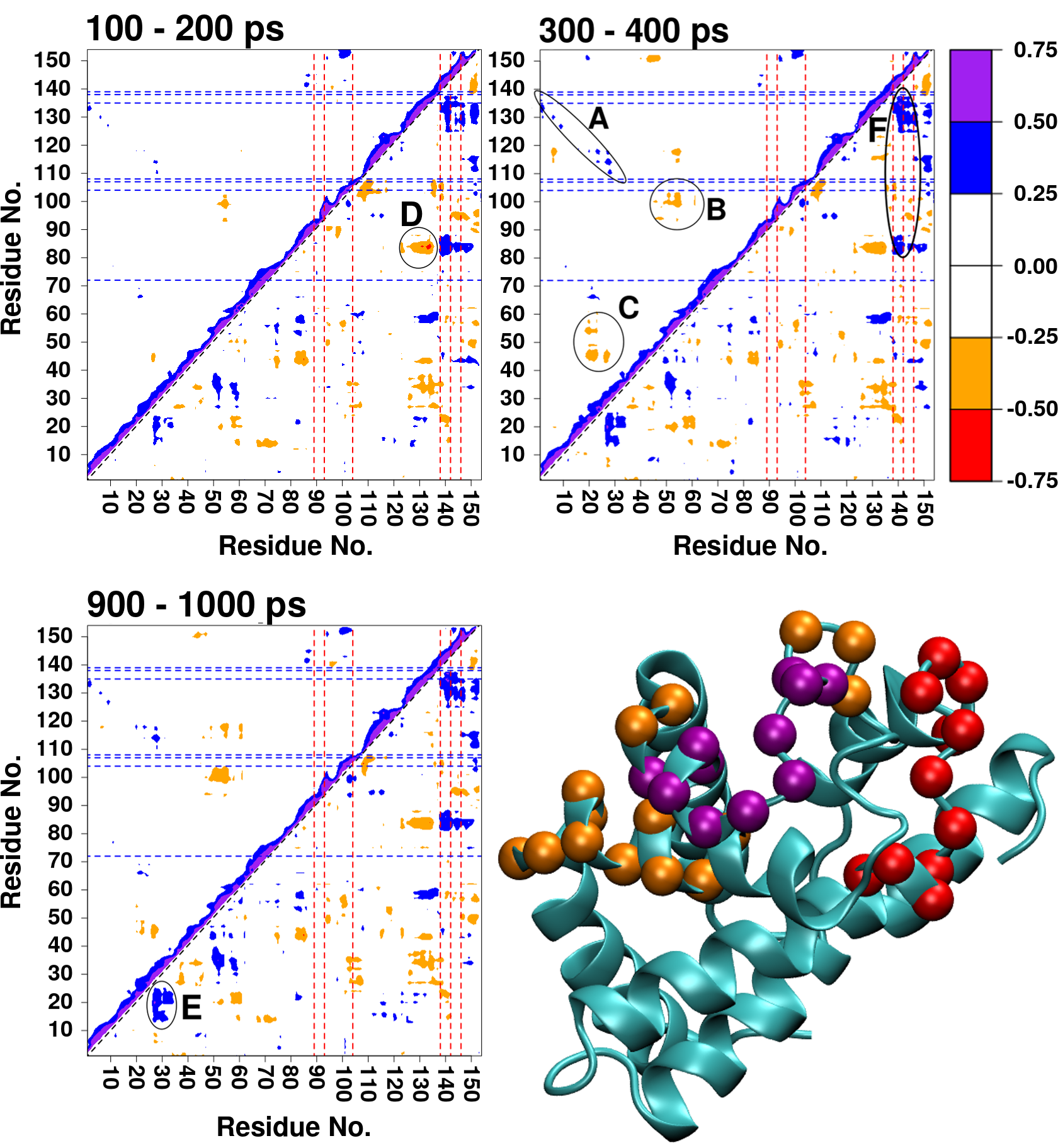}
\caption{The DCCM from the 1 ns free dynamics simulations in which the
  Xe2$\rightarrow$Xe1 transition is observed for 0101. The DCCM was
  generated at 100 ps intervals. The Upper matrix shows DCCM for
  Xe2$\rightarrow$Xe1 transition whereas the lower matrix is
  $\Delta$DCCM for a given time interval and the 0000 reference
  state. DCCMs correspond to transition times between 100 to 200 ps,
  300 to 400 ps, and 900 to 1000 ps. Residue 154 is the Xe
  atom. Vertical red lines indicate the residues that form the Xe1
  whereas horizontal blue lines indicate the residues that form the
  Xe2. Particular features in the maps are labeled from A to F with
  circles around them. Features B and C of DCCM visualized on
  myoglobin structure for Xe2$\rightarrow$Xe1 in 0101. The color code
  is as follows: C$_{\alpha}$ of feature B (Ala19 to Leu29 and Phe43
  to Ser58) orange, C$_{\alpha}$ of feature C (Phe46 to Lys56 and
  Ala94 to Leu104) is red and C$_{\alpha}$ that are part of both
  features is purple. The blue line emphasizes the correlated motions
  between given atoms.}
\label{fig6}
\end{center}
\end{figure}

\noindent
It is also of interest to analyze DCCMs as a function of simulation
time which was done for the unbiased simulations for the
Xe2$\rightarrow$Xe1 transition (0101 occupation), see
Figure~\ref{fig6}. The DCCMs were generated in 100 ps intervals by
pooling and averaging over trajectories for which Xe migration
occurred within the corresponding time interval. Hence, for 300 - 400
ps (upper right panel figure \ref{fig6}) all trajectories for which Xe
migration occurred in this time window were analyzed
together. Difference DCCMs ($\Delta$DCCMs) show the differences
between 0101 and the reference 0000 state. The upper left matrix shows
DCCM for Xe2$\rightarrow$Xe1 transition whereas the lower right matrix
is $\Delta$DCCM for a given time interval and with 0000 as the
reference state. For the DCCMs, three common features have been
observed for all time intervals with varying intensities, A)
Correlated motions between residues Val1 to Ile30 and 110 to 135 which
are long-range interaction. B) Anti-correlated motions between
residues Arg45 to Lys62 and Ser92 to Tyr103 which are mid-range
interactions, and C) Anti-correlated motions between residues Asp20 to
Arg31 and Glu41 to Asp60. These features are labelled in
Figure~\ref{fig6}.\\

\noindent
$\Delta$DCCMs show the structural effects induced by the Xe atom on
the local structure around the Xe1 and Xe2 pockets and the global
structure of the protein. Upon transition, correlated motions in
myoglobin are significantly altered. Compared with the empty protein,
strong correlated (feature D, $C_{ij} \geq 0.75$, see
Figure~\ref{fig6}) and anti-correlated (feature E, $C_{ij} \leq
-0.75$, see Figure~\ref{fig6}) motions are observed. No difference in
correlation motions ($C_{ij} \geq 0.25$) been observed from N-terminus
to Asp27. Further, Phe138, Ile142, Tyr146 of Xe1 and Leu135, Phe138,
Arg139 of Xe2 show correlated motions in 0101 compared with the empty
protein. In particular, residues Phe138, Ile142, Tyr146 of pocket Xe1
show correlated motions with residues Ala125 to Arg139 and His82 to
Lys87 when considering the $\Delta$DCCM maps. Strongly correlated
motions involving residues Phe138 to the C-terminus were also found
(see Figure~\ref{fig6}, label F)\\

\noindent
Overall DCCM and $\Delta$DCCM show that correlated motions of the
protein are subject to transition direction and neighboring pocket
occupation upon Xe transition, as shown in Figure~\ref{fig5}. The
correlation/anti-correlation in residue motion accompanying transition
of Xe between neighboring pockets and depending on occupation state
are indicative of allostery. The results for the 0101 occupation
suggest that insertion of xenon into the protein significantly
decreased the correlated/anti-correlated motions compared with the
empty protein.\\

\section{Discussion}
The present work determined the energetics for Xe migration pathways
between neighboring docking sites in Mb from biased and unbiased MD
simulations for different occupation states of the protein. Generally
it was found that differential stabilization of the endpoints and
barrier heights for Xe in the initial and final state can depend on
both, occupation of the remaining docking sites and the direction in
which the transition was scanned. Unbiased simulations for one
transition confirmed that barrier heights from biased simulations
(here umbrella sampling) are representative. However, the distribution
of transition times also showed that associating a single $\Delta G$
value to a given transition is not particularly meaningful from repeat
US simulations for one transition and from the finding that transition
times are distributed over 1 ns from unbiased simulations on the same
time scale. In the following, a structural interpretation for specific
transition paths is attempted and the results obtained are put in a
broader context.\\

\begin{figure}[H]
\begin{center}
\includegraphics[width=0.3\linewidth]{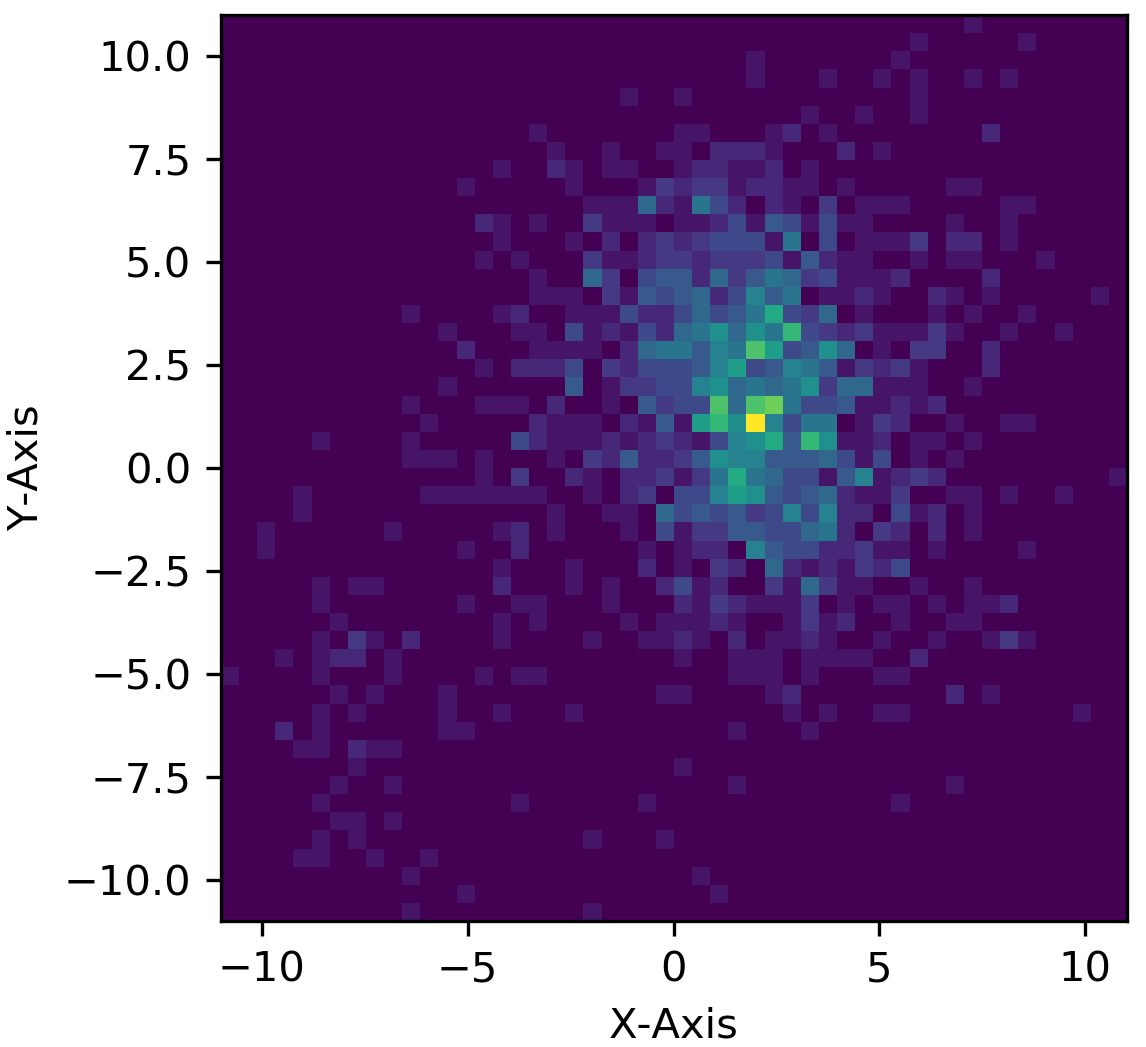}
\includegraphics[width=0.3\linewidth]{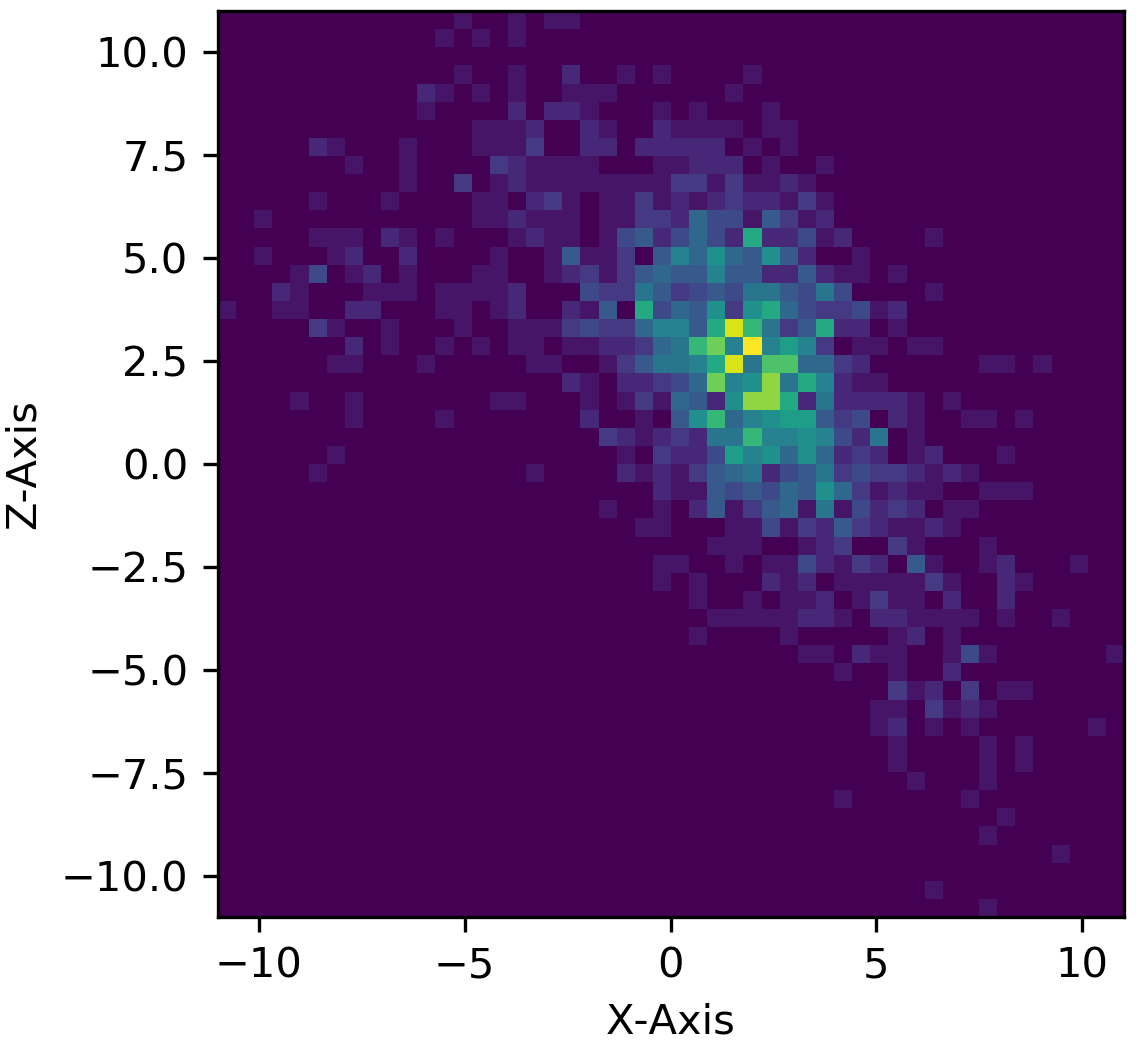}
\includegraphics[width=0.3\linewidth]{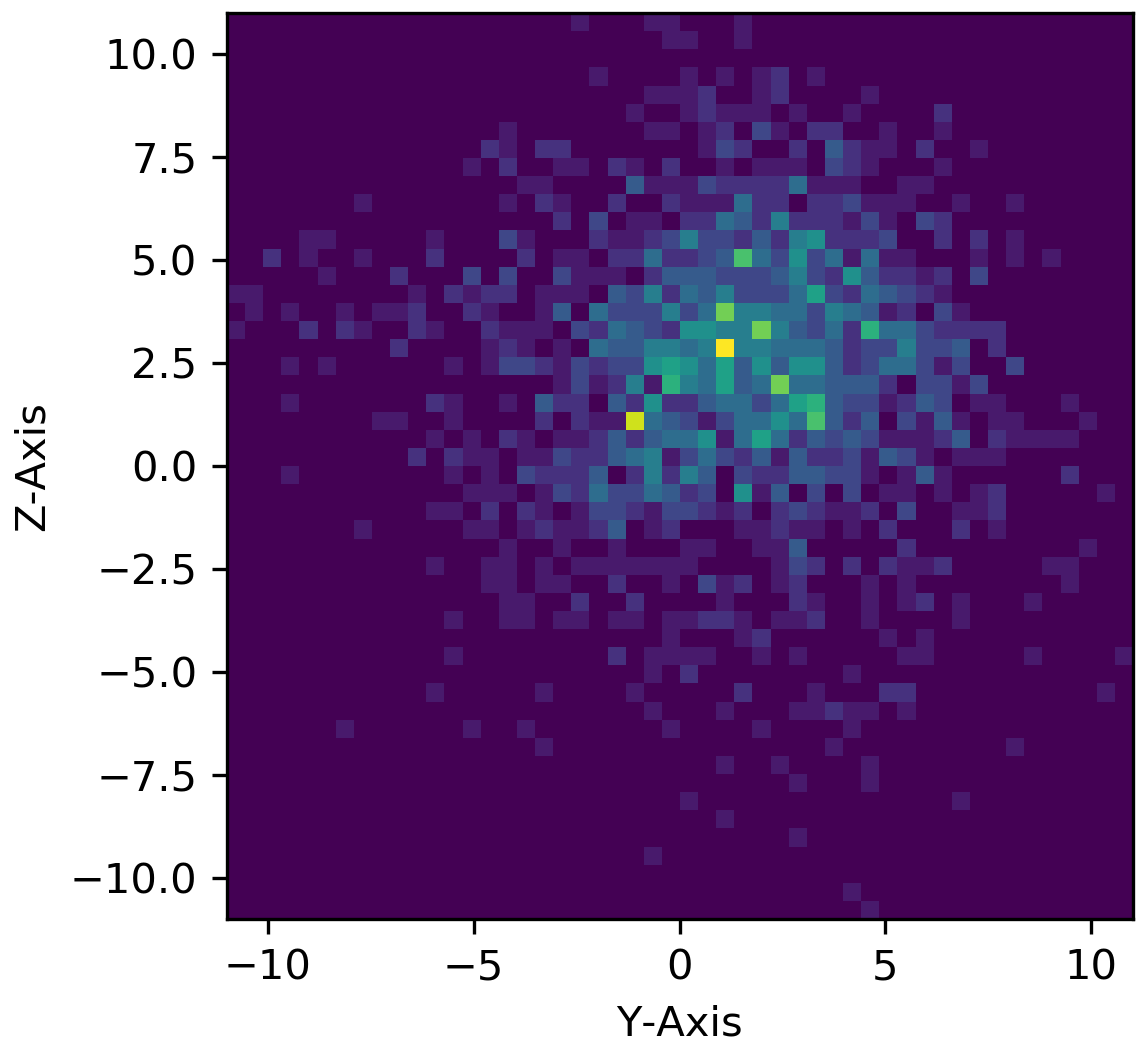}
\includegraphics[width=0.3\linewidth]{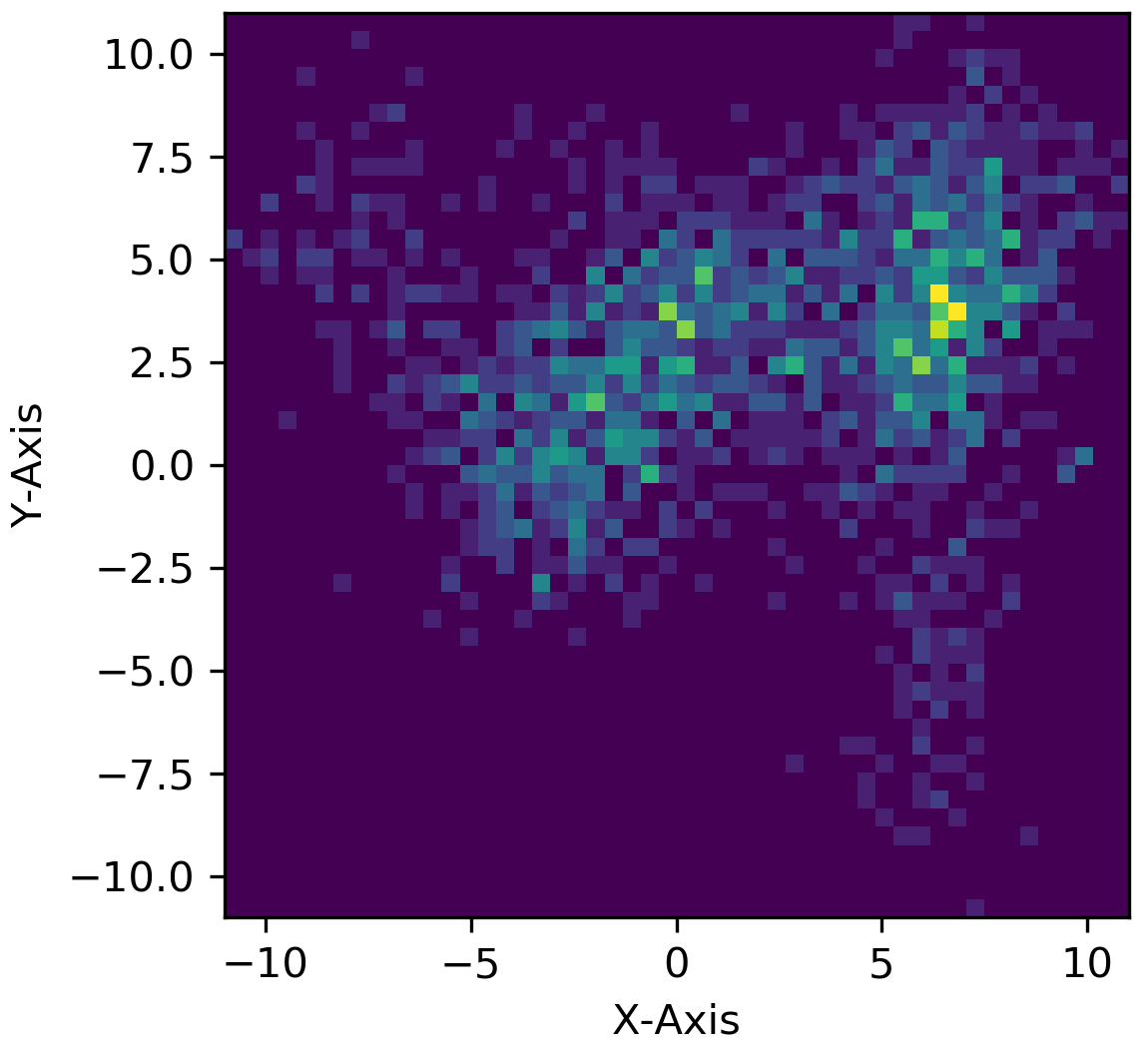}
\includegraphics[width=0.3\linewidth]{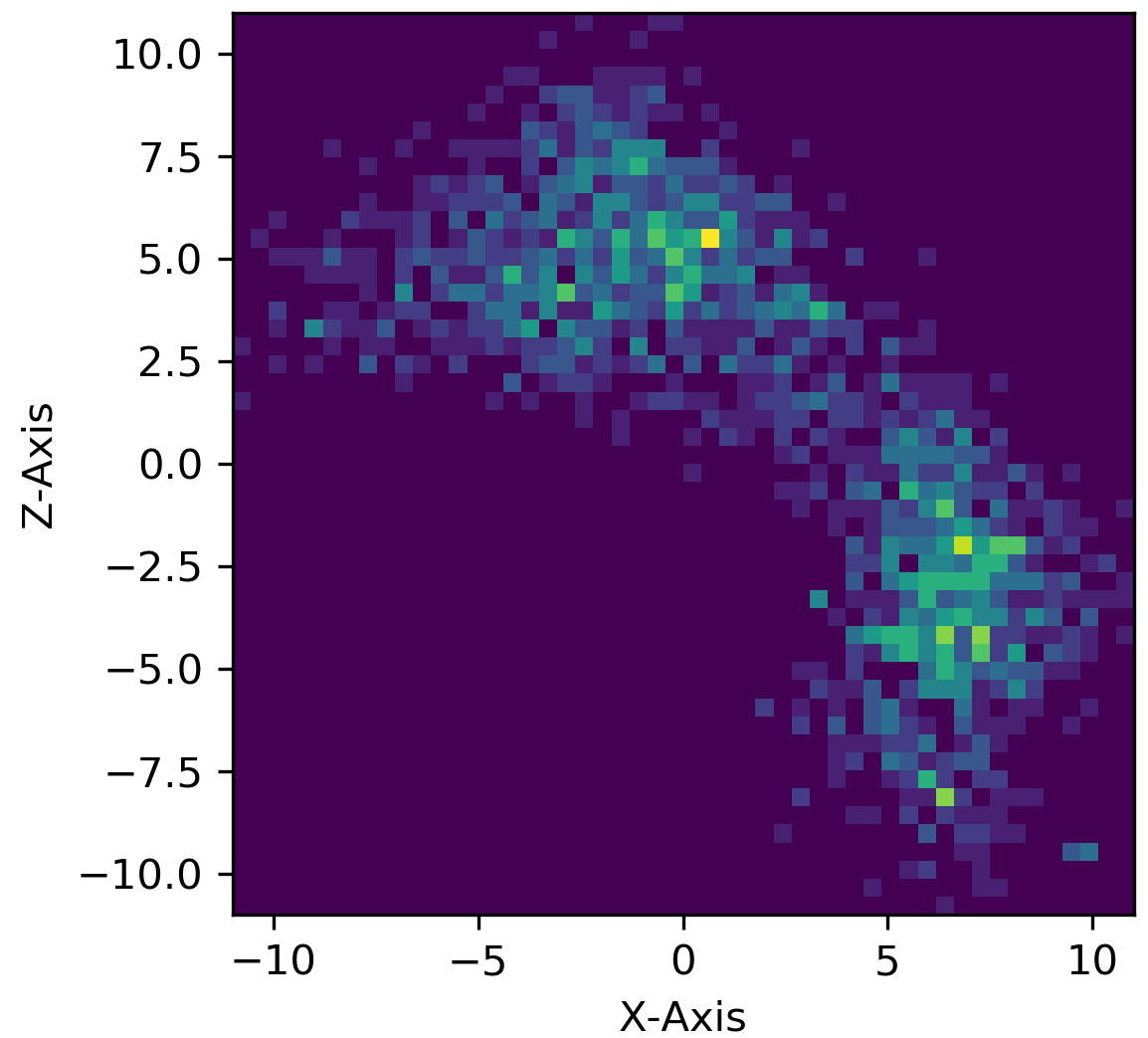}
\includegraphics[width=0.3\linewidth]{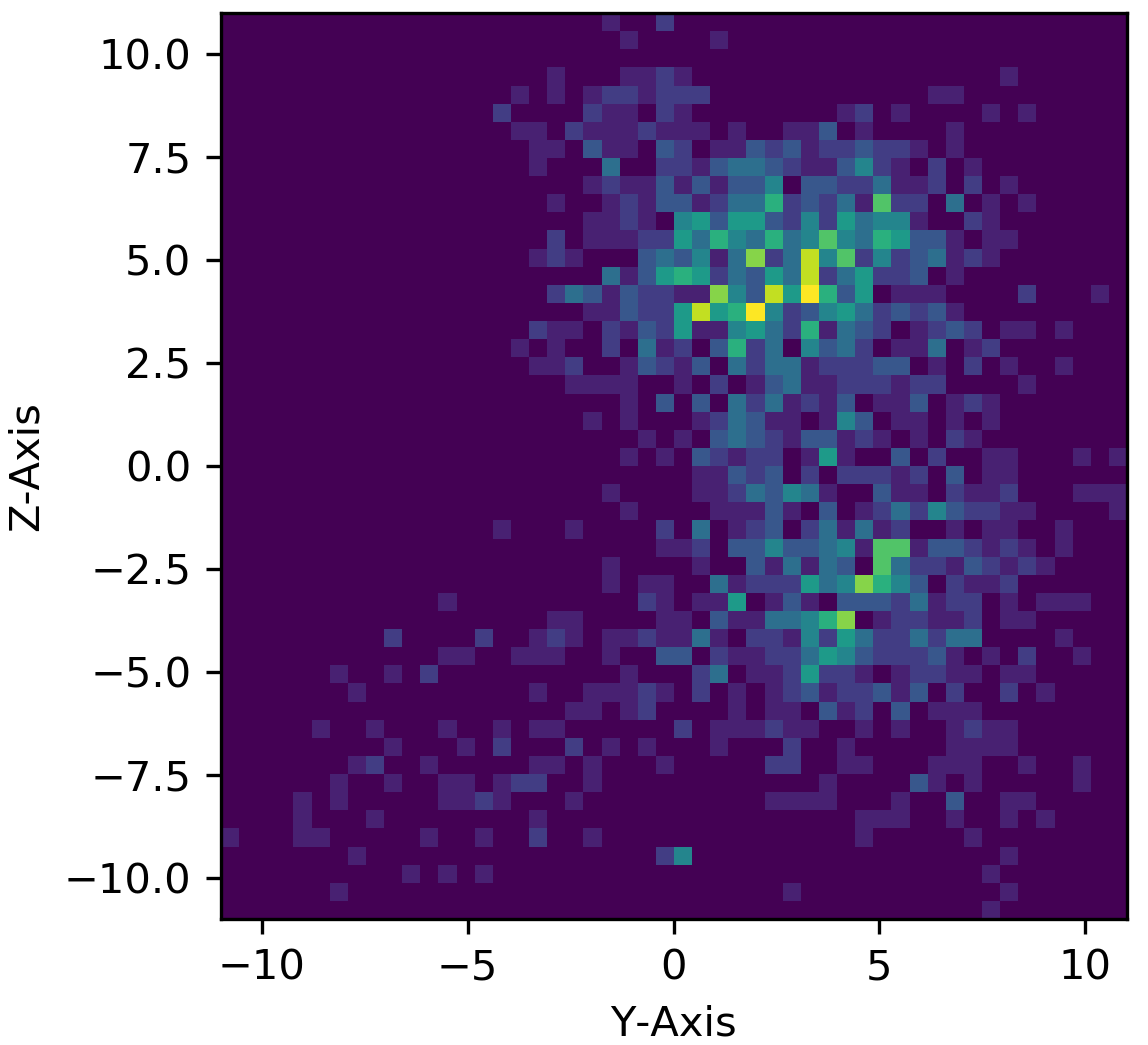}
\includegraphics[width=0.45\linewidth]{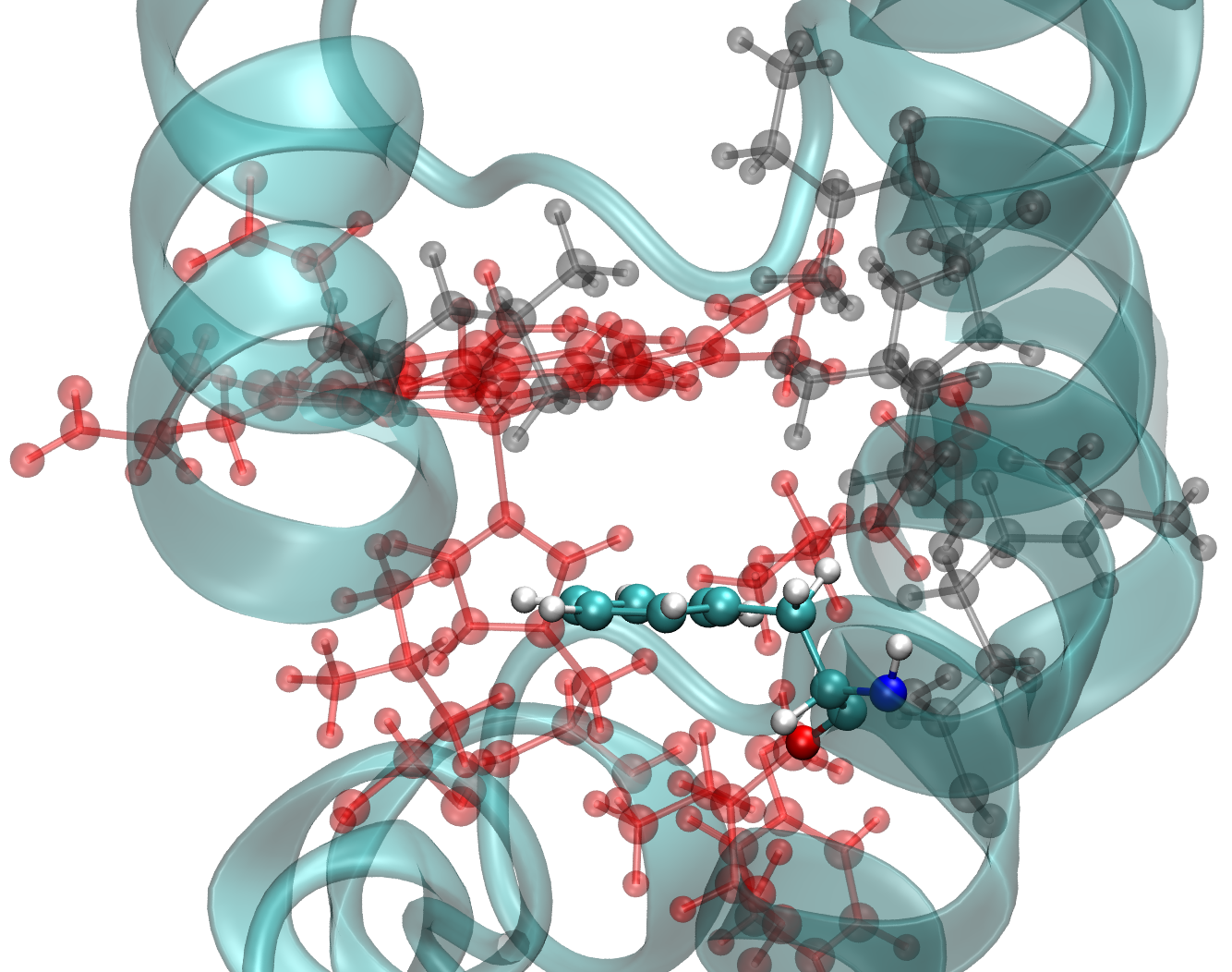}
\includegraphics[width=0.45\linewidth]{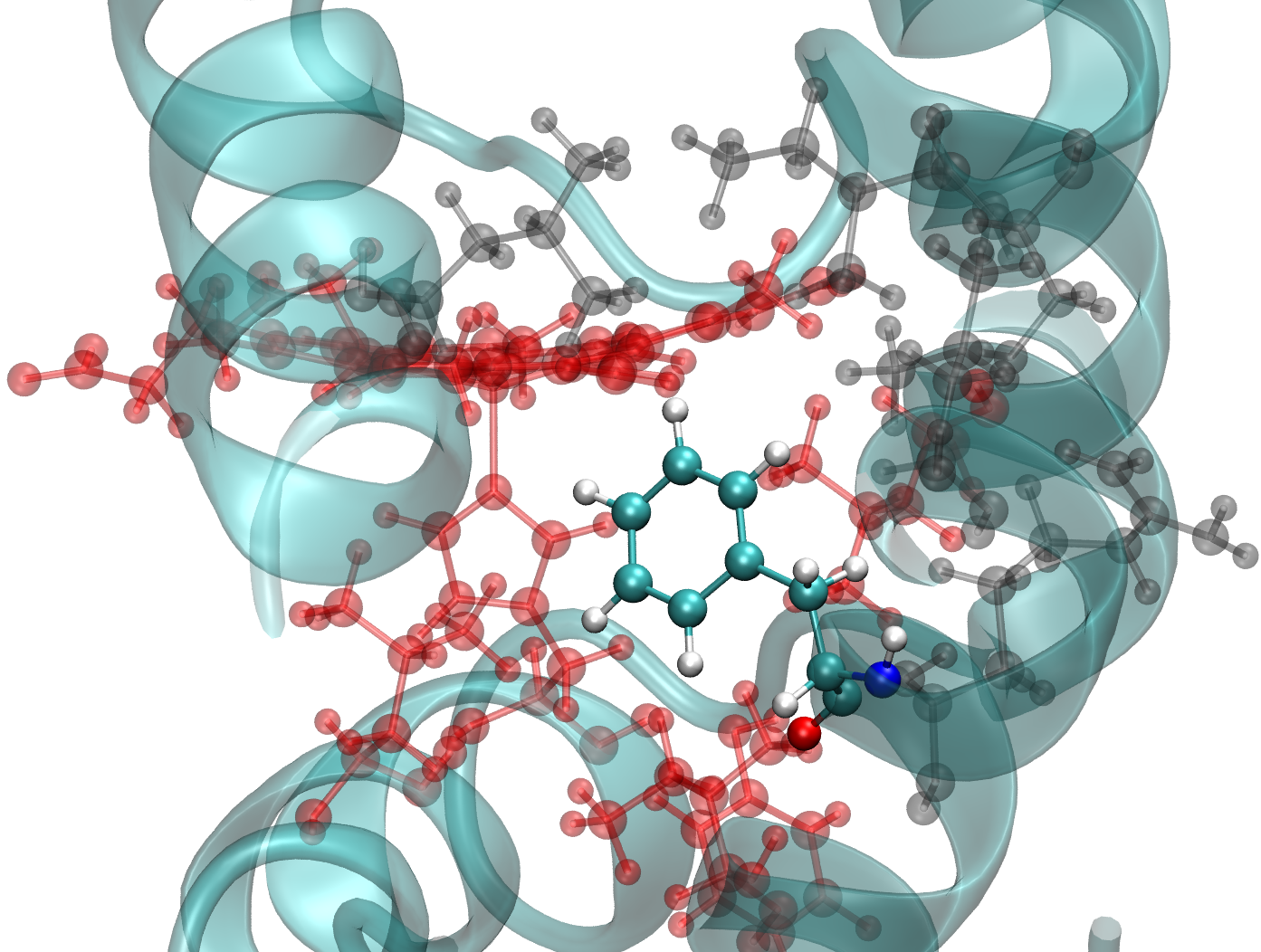}
\caption{The center of mass position of the Phe138 side chain
  projected onto the $xy$, $xz$, and $yz$ planes for the
  Xe2$\rightarrow$Xe1 transition in 0101 (Top) Xe1$\rightarrow$Xe2
  transition 1001 (Middle). The data was extracted from the umbrella
  sampling simulations at the TS and for the two windows preceding and
  following the TS. The lighter color means higher probability and the
  two maxima shown in the middle row correspond to the open and close
  conformations of Phe138. In the bottom row the open (left) and
  closed (right) passage with residue Phe138 as the gate is shown. The
  red residues are part of pocket Xe1 (front) whereas black residues
  are part of pocket Xe2 (back).}
\label{fig7}
\end{center}
\end{figure}

\noindent
{\it Structural interpretation for Xe2$\rightarrow$Xe1 (0101) and for
  Xe1$\rightarrow$Xe2 (1001):} The activation barrier was lowered by 3
kcal/mol in the reverse transition of Xe1$\leftrightarrow$Xe2 in 0101
compared to the forward transition in 1001, see Figure~\ref{fig2}. The
side-chains are investigated to understand the decrease in barrier
heights as a function of local motions. Figure~\ref{fig7} shows the
CoM position of the Phe138 side chain is analyzed on XY, XZ, and YZ
planes for Xe2$\rightarrow$Xe1 transition in 0101 (Top)
Xe1$\rightarrow$Xe2 transition 1001 (Bottom).\\

\noindent
The results show that the side chain of Phe138 has two conformations
(open and closed) for 1001 whereas it remains in the open conformation
for 0101. The conformation in 0101 allows easier access for Xe to the
Xe1 pocket, whereas switching between the open and closed
conformations through side-chain rotation inhibits Xe mobility between
the pockets in 1001. One should note that sampling distinct
conformations based on the occupation state may be linked to
allostery. Overall, Phe138 acts as a gate between Xe1 and Xe2
pockets. Features A, B, and C of the DCCM for the trajectories with
transition times of 300 ps to 400 ps are shown in Figure
\ref{fig6}. These features report on both neighboring and
through-space correlations. Transition times of 300 to 400 ps
correspond to $\sim 3$ kcal/mol activation barrier which is on-par
with calculated 1D PMF for the given transition. \\

\noindent
The forward activation barrier for Xe1$\leftrightarrow$Xe2 in 1010 was
5 kcal/mol higher compared with the reverse barrier and the free the
energy of the state with Xe in pocket Xe1 depends on whether the
transition is followed in the forward (black) or the reverse (red)
direction, see Figure~\ref{fig2}C.  For a structural interpretation,
the averaged protein structures of the first umbrella for the forward
direction, and that for the last umbrella for the backward direction
are considered, whereby both correspond to $r_{\rm c}$ = 0. The
C$_{\alpha}$ atoms of the residues forming the Xe1 pocket (see Table
\ref{tab:pockets}) are superimposed. The root mean squared deviations
(RMSD) of all heavy atoms of the residues forming the Xe1 pocket are
collected in Table S2. Most importantly, the RMSD for all
heavy atoms forming the Xe1 pocket was 1.33 \AA\/, which shows that
there are substantial structural differences even for the few residues
forming the Xe1 pocket depending on whether the transition was scanned
in forward or in reverse direction. The largest RMSD computed for all
heavy atoms arises from Leu104 and Phe138 residues.\\

\noindent
{\it Correlated Motions Depending on Transition Time:} The total
correlation of a group of residues can be used to compare the
strengths of dynamical coupling depending on the occupation state or
depending on a particular time window of the simulations. For this,
the correlation coefficients $C_{ij}$ involving groups of residues $i$
to $j$ were added to determine $c_{ij} =1/N \sum_{i=1}^{n} \sum_{j =
  1}^{m} C_{ij}$, where $n$ and $m$ are the numbers of residues over
which a given feature extends, $N$ is the total number correlation
coefficient ($N = m \times n$). As an example for Feature B,
correlation coefficients of Ala19 with Phe43 to Ser58 are summed; then
coefficients of Asp20 with Phe43 to Ser58 are added to the previous
summation; summations are continued until the coefficients of Leu29
with Phe43 to Ser58 added. Finally, total number is divided by
$N$. For features B and C, $c_{ij}$ presented as a function of time
and reported in Figure S1. All $c_{ij}$ between Ala19 to
Leu29 and Phe43 to Ser58 for B, and Phe46 to Lys56 and Ala94 to Leu104
are summed and divided by the total number of residues for each
feature to determine how the total correlation for a group of residues
changes as a function of simulation time. For feature B there was an
initial increase then it was stable until 0.8 ns, afterwards increase
in magnitude to the initial values levels is observed.  For feature C
the total anti-correlation increased until 0.4 ns after which it
stabilizes. Hence, depending on the group of residues considered the
total correlation between participating residues can also behave
differently as a function of simulation time.\\

\begin{figure}[H]
\begin{center}
\includegraphics[width=0.7\linewidth]{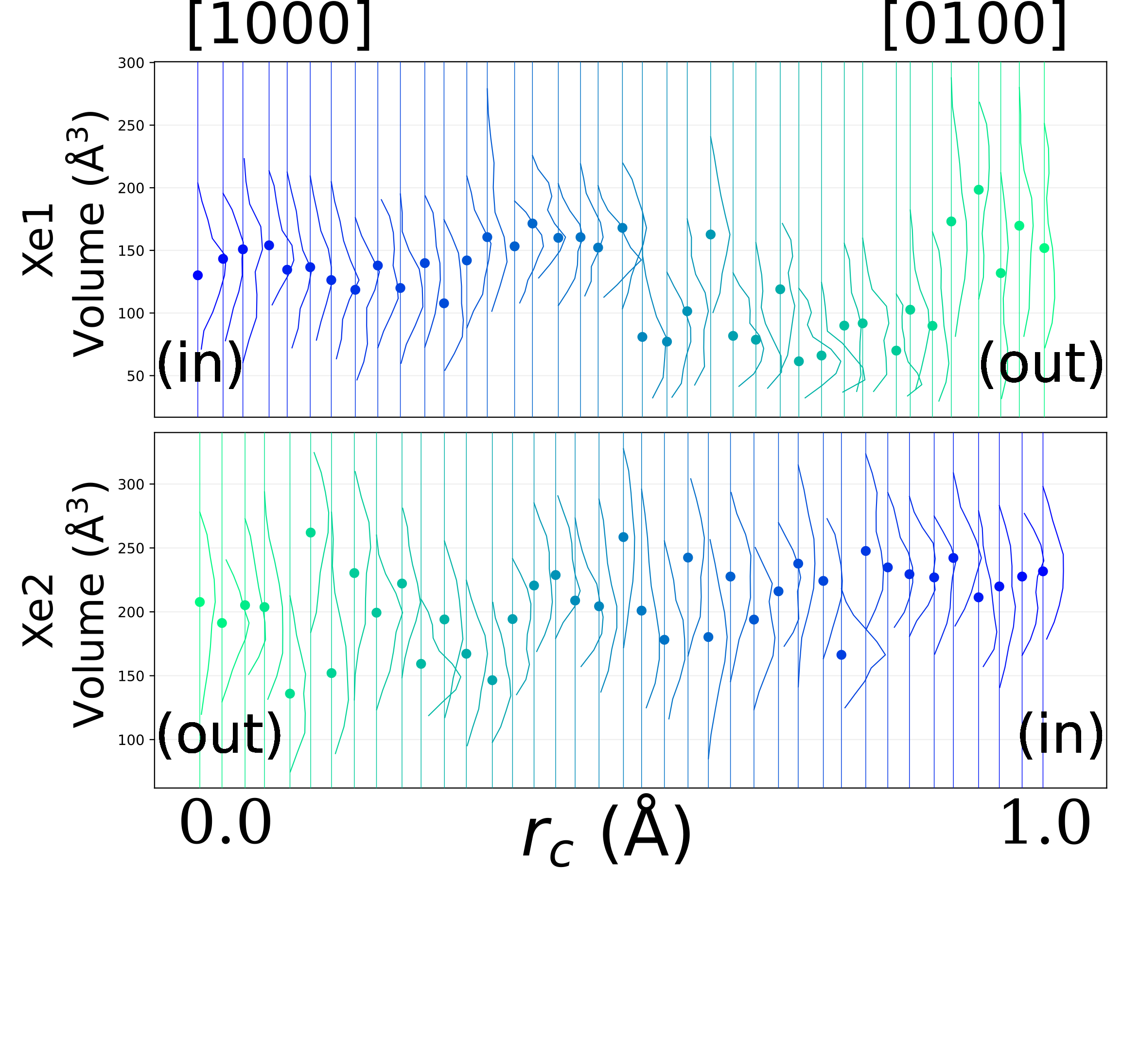}
\caption{The volume of Xe1 and Xe2 for the Xe1$\rightarrow$Xe2
  transition starting for the 1000 state. (In) and (Out) refer to the
  Xe atom inside the initial (Xe1) pocket and the final (Xe2) pocket,
  respectively. The points are the median of the pocket volumes.}
\label{fig8}
\end{center}
\end{figure}

\noindent
{\it Analysis of pocket volumes:} Another property that provides
information about internal protein rearrangements is the volume of the
internal cavities and their fluctuations along the ligand migration
pathway. This was done for the Xe1 and Xe2 volumes along the
Xe1$\leftrightarrow$Xe2 transition pathway. The distribution of pocket volumes
from the US simulations for Xe1$\rightarrow$Xe2 as a function of
reaction coordinate is shown in Figure~\ref{fig8}. The filled circles
are the median pocket volumes along the reaction coordinate,
determined using the 50th percentile of the cumulative probability
distribution, based on a kernel density estimation.\\

\noindent
The distributions and the median volumes demonstrate that depending on
the position of the xenon atom along the pathway the volumes of the
pockets respond. The volume distributions are non-Gaussian, sometimes
with more than one peak, and vary from one umbrella window to the
next for each pocket considered. The average volume of the Xe1 pocket
(top panel) first varies between 100 and 166\,\AA\/${^3}$ and sharply
decreases by $\sim 50$\,\% after $r_{\rm c} > 0.5$ \AA\/, when Xe
leaves the Xe1 pocket, until $r_{\rm c} = 0.8$ \AA\/. After that, the
Xe1 pocket volume returns to its initial value which was $\sim 150$
\AA\/${^3}$. For the Xe2 pocket the average volume is $\sim 200$
\AA\/${^3}$ but along the transition pathway the fluctuation around
this value is $\sim\pm 50$\,\AA\/${^3}$. The volume of the Xe2 pocket
decreases for the first part of the transition (green in Figure
\ref{fig8}. The Xe2 pocket volume fluctuates more in $0.5 < r_{\rm c}
> 0.6$ \AA\/ region which corresponds to the ``transition" region for
Xe1$\leftrightarrow$Xe2.\\

\noindent
{\it The DP$\leftrightarrow$Xe4 Transition:} Physiologically, the DP
to Xe4 transition is most relevant.\cite{Scott2001,olson:2007} The
findings from repeat forward and reverse simulations (see Figure
\ref{fig4}) indicate that Xe in DP is higher in energy compared with
Xe in Xe4 by 4.22 kcal/mol on average. The free energy barrier
separating the two states is 3.79 kcal/mol in the forward and 3.62
kcal/mol in the reverse direction on average. The differential
stabilization in favour of Xe in Xe4 is consistent with flash
photolysis experiments on CO in Mb which find spontaneous CO diffusion
from docking site ``B'' to Xe4 on the nanosecond time
scale.\cite{Schotte2003a} Similarly, implicit ligand sampling
simulations also found stabilization of Xe in Xe4 to be more
favourable than in DP.\cite{Cohen2006} These comparisons provide
independent validation of the present simulations.\\

\begin{figure}[H]
\begin{center}
\includegraphics[width=0.9\linewidth]{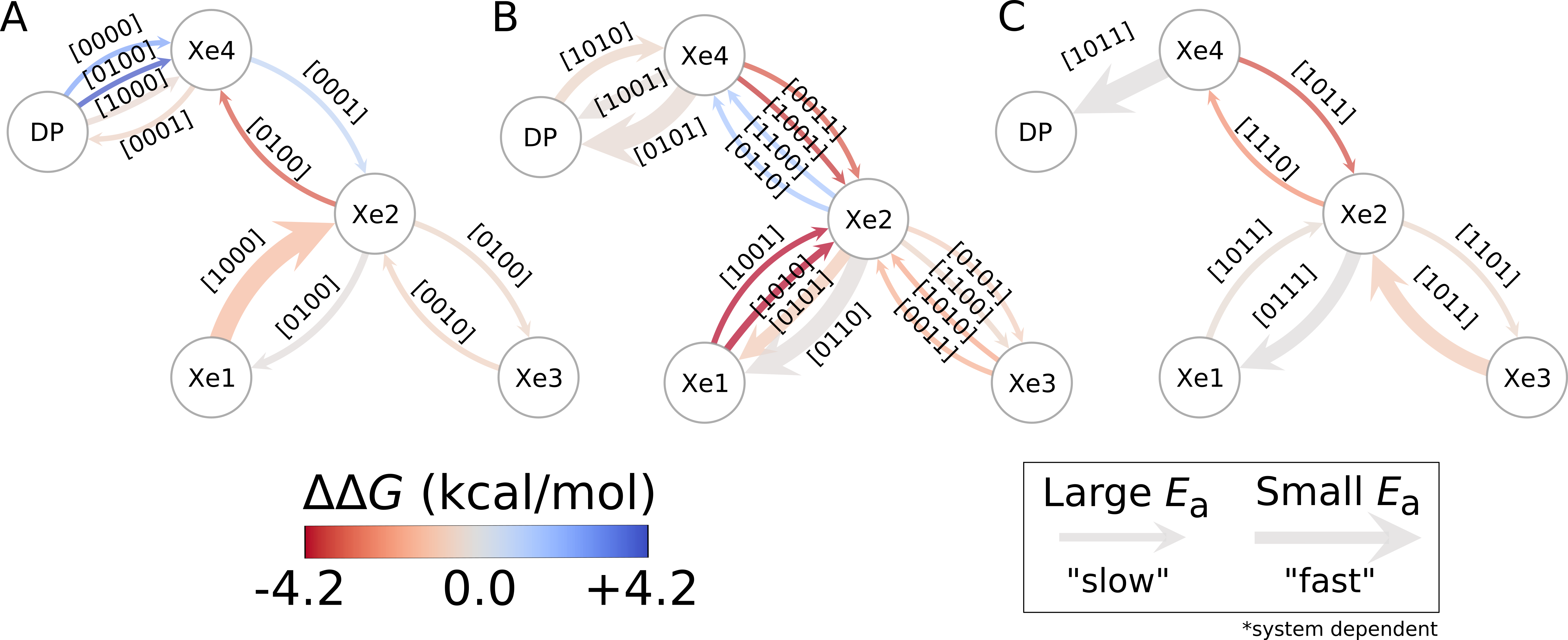}
\caption{Network representation illustrating the transitions studied
  for systems with (A) one xenon atom, (B) two xenon atoms and (C)
  three xenon atoms occupying the protein.  Arrows are colored by the
  difference in free energy between the initial and the final states
  and labelled by the initial occupation state (i.e. 1000 for
  Xe1$\rightarrow$Xe2).  The size of the arrows is proportional to the
  ratio of rate constant for transitions, estimated using the
  Arrhenius equation at 300 K, assuming the same Arrhenius prefactor
  for all transitions, and can be compared between each of the three
  distinct occupancies.}
\label{fig9}
\end{center}
\end{figure}

\noindent
{\it State dependence on xenon transitions in Mb:} Specific
occupancies of internal pockets in myoglobin introduce correlation
between distant sites of the protein. Network representations, as
shown in Figure \ref{fig9}, provide an overarching view of all
migration pathways considered.  Nodes indicate the identity of the
pockets as their associated residues are known through experiment. In
such a directed graph, edges represent transitions which were analysed
using their respective potential of mean forces to provide estimates
of relative stabilization energies and barriers, see Table S1. For xenon migration in myoglobin, the occupied versus
empty state of a particular pocket has a measurable impact on the
kinetic and thermodynamic properties of the transitions, as seen in
Figure \ref{fig9}. Transitions which require excess free energy to
occur when only one xenon atom occupies the protein (Figure
\ref{fig9}A), in DP to Xe4, and Xe4 to Xe2, can occur spontaneously
when an additional xenon atom occupies either Xe1 or Xe3, respectively
(Figure \ref{fig9}B). Hence, occupation of specific internal pockets
can modulate energy barriers for a given transition path which also
leads to changes in the relative kinetic rates. For instance, the
forward transition between Xe1 to Xe2 is more kinetically favoured
when internal pockets are unoccupied (a factor of $\sim 10$ times
faster than the reverse transition) and are slower in proportion to
comparable transitions when an additional xenon atom is present in a
distant pocket (approximately a factor of 20 times slower than the
reverse transition, (Figure \ref{fig9}B)). The same trend is observed
for this transition when 3 xenon atoms are occupying the protein
(Figure \ref{fig9}C), where the reverse transition is also
favoured. \\

\section{Conclusion}
In conclusion, both energetic and structural analyses were performed
to better understand the molecular level characteristics of the ligand
movements inside myoglobin and whether/how they relate to
allostery. These analyses are further supported by the DCCMs,
side-chain, and volume calculations. For a particular transition, the
occupation of the neighboring pockets greatly affects both energetics
and structural dynamics of the protein. The activation barriers and
the relative stabilizations for the ligand in the initial and final
pocket of a transition are distributed, rather than a single, discrete
value, that depends on the sampled initial meta-stable
state. Side-chain degrees of freedom were found to gate particular
transition pathways, as is the case of Phe138. The dependence of
barrier heights on occupation state and scan-direction suggest that
non-local communication between different protein sites occurs which
is a hallmark of allosteric control. The DCCMs, which characterize the
correlated motions with the protein further support this
interpretation. Finally, a network representation for all ligand
migration pathways considered highlights the state dependence of
thermodynamic and kinetic properties. The sum of all these
observations, which are based on a molecular-level characterization of
xenon migration and protein dynamics, indicate that Mb is an
allosteric protein. This finding is also consistent with earlier work
on Mb which proposed that Mb is an allosteric enzyme by pointing to
the important role of the internal cavities to speed up bimolecular
reactions by increasing the concentration of the physiological ligands
and by reducing or entirely eliminating competing reaction
pathways.\cite{frauenfelder:2001}\\

\section{Acknowledgments}
The authors gratefully acknowledge financial support from the Swiss
National Science Foundation through grant 200021-117810 and to the
NCCR-MUST.  \\

\bibliography{references}

\end{document}

% \date{\today}

\begin{table}[]
\caption{The activation energies ($E_a$) and thermodynamic parameter
  ($\Delta \Delta G$) from Figure 2 for forward and reverse
  transitions, for each system where $N_{\mathrm{Xe}}$ is the number
  of xenon atoms inside the protein. $\Delta G_{\rm P_1}$ and $\Delta
  G_{\rm P_2}$ represent the free energy for the starting and terminal
  pocket during the transition, respectively. Units for free energies
  are in kcal/mol.}  \centering
\begin{tabular}{|c|l|l|l|c|c|l|c|}
\hline
$N_{\mathrm{Xe}}$ & State &    P$_1$ & P$_2$ &    $\Delta G_{\rm P_1}$ &    $\Delta G_{\rm P_2}$ &                    $E_a$ &                     $\Delta\Delta G$ \\
\hline
        1 &  0000 &   DP &  Xe4 &    0.0 &  2.4 &               5.7 &                -2.4  \\
        1 &  0001 &  Xe4 &  Xe2 &    0.0 &  0.8 &               8.9 &                -0.8  \\
        1 &  0001 &  Xe4 &   DP &    0.6 &    0.0 &               3.2 &                 0.6  \\
        1 &  0010 &  Xe3 &  Xe2 &    0.6 &    0.0 &               2.5 &                 0.6  \\
        1 &  0100 &  Xe2 &  Xe1 &  0.1 &    0.0 &               1.8 &                 0.1 \\
        1 &  0100 &  Xe2 &  Xe3 &  0.5 &    0.0 &               2.5  &                 0.5 \\
        1 &  0100 &  Xe2 &  Xe4 &  3.4 &    0.0 &               4.5  &                 3.4 \\
        1 &  0100 &   DP &  Xe4 &    0.0 &   4.1 &                5.4 &                 -4.1  \\
        1 &  1000 &  Xe1 &  Xe2 &  1.4 &    0.0 &               0.6 &                 1.4 \\
        1 &  1000 &   DP &  Xe4 &  0.3 &    0.0 &               2.1 &                 0.3 \\ \hline
        2 &  0011 &  Xe3 &  Xe2 &  1.6 &    0.0 &               4.4 &                 1.6  \\
        2 &  0011 &  Xe4 &  Xe2 &  3.4 &    0.0 &               4.5  &                 3.4 \\
        2 &  0101 &  Xe2 &  Xe1 &  0.9 &    0.0 &               1.6  &                 0.9 \\
        2 &  0101 &  Xe2 &  Xe3 &  0.8 &    0.0 &               4.5  &                 0.8 \\
        2 &  0101 &  Xe4 &   DP &  0.3 &    0.0 &               1.2  &                 0.3  \\
        2 &  0110 &  Xe2 &  Xe1 &  0.1 &    0.0 &               1.3  &                 0.1  \\
        2 &  0110 &  Xe2 &  Xe4 &    0.0 &  1.5 &               9.3 &                -1.5 \\
        2 &  1001 &  Xe1 &  Xe2 &  4.2 &    0.0 &               3.0  &                 4.2  \\
        2 &  1001 &  Xe4 &  Xe2 &  3.9 &    0.0 &               4.7  &                 3.9 \\
        2 &  1001 &  Xe4 &   DP &  0.2 &    0.0 &               1.8  &                 0.2 \\
        2 &  1010 &  Xe1 &  Xe2 &  4.2 &    0.0 &               2.4  &                 4.2 \\
        2 &  1010 &  Xe3 &  Xe2 &  1.8 &    0.0 &               4.1  &                 1.8 \\
        2 &  1010 &   DP &  Xe4 &  0.5  &    0.0 &                1.9 &                 0.5 \\
        2 &  1100 &  Xe2 &  Xe3 &  0.5 &    0.0 &               3.3 &                 0.5 \\
        2 &  1100 &  Xe2 &  Xe4 &    0.0 &  1.5 &               9.3 &                -1.5 \\ \hline
        3 &  0111 &  Xe2 &  Xe1 &  0.1 &  0.1 &               1.4 &                 0.0 \\
        3 &  1011 &  Xe1 &  Xe2 &  0.2  &    0.0 &                2.5 &                  0.2  \\
        3 &  1011 &  Xe3 &  Xe2 &  0.8  &    0.0 &               1.3 &                 0.9 \\
        3 &  1011 & Xe4 &  Xe2 &  3.5 &    0.0 &               5.7 &                 3.5 \\
        3 &  1011 &  Xe4 &   DP &  0.0 &    0.0 &               1.1  &                 0.0 \\
        3 &  1101 &  Xe2 &  Xe3 &  0.4 &    0.0 &               3.4 &                 0.4 \\
        3 &  1110 &  Xe2 &  Xe4 &  2.3 &    0.0 &               7.3 &                 2.3 \\\hline
\end{tabular}
    \label{sitab1}
\end{table}

\begin{figure}[H]
\begin{center}
\includegraphics[width=0.75\linewidth]{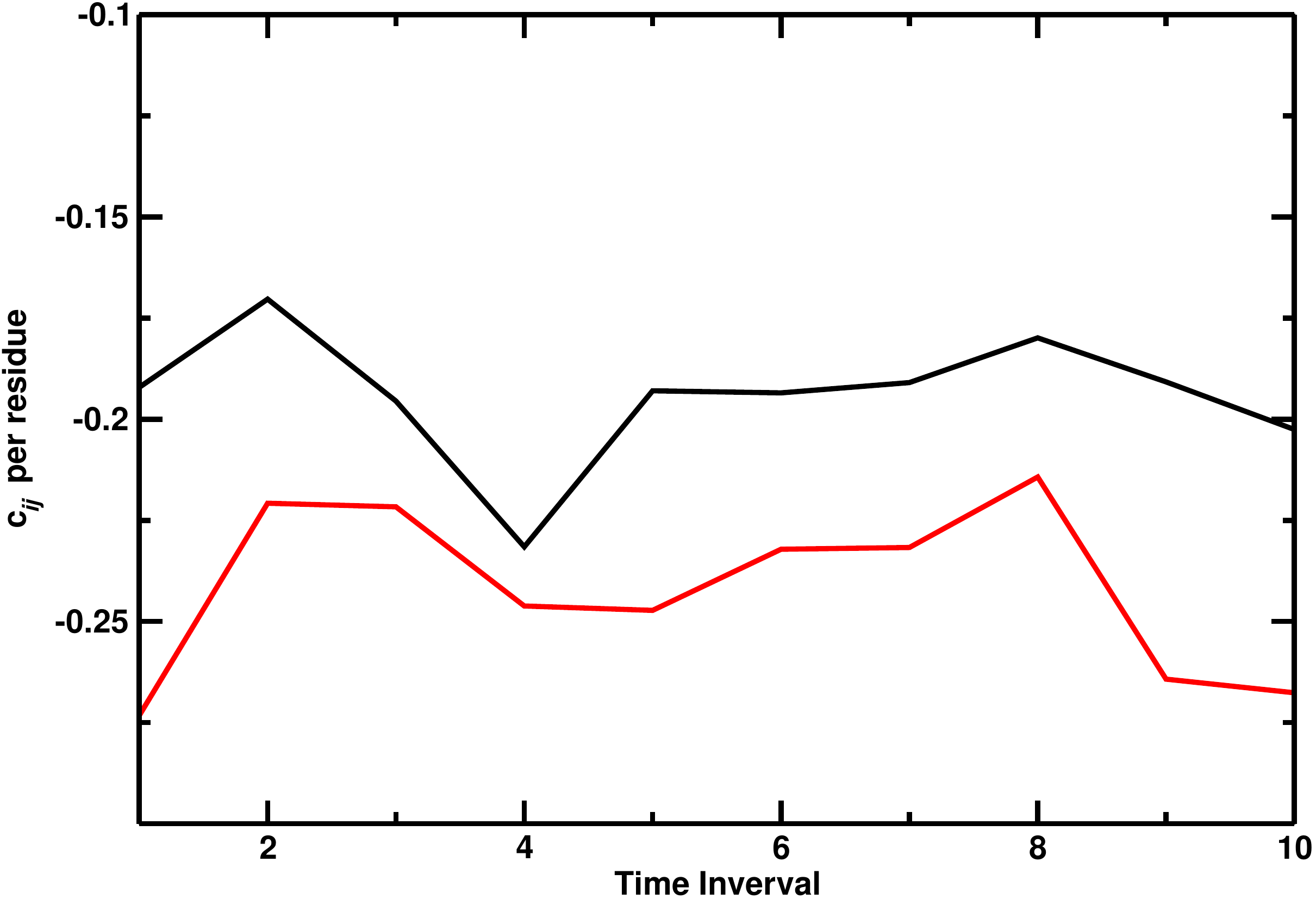}
\caption{Cumulative correlation coefficients $c_{ij}$ per residue of B
  (black) and C (red) features from DCCMs for Xe2$\rightarrow$Xe1 in
  0101. The B feature consists of the residues of Ala19 to Leu29 and
  Phe43 to Ser58, The C feature consists of the residues of Phe46 to
  Lys56 and Ala94 to Leu104.}
\label{sifig1}
\end{center}
\end{figure}

\begin{table}[]
\caption{The root mean square displacement (RMSD) in \AA\/ for Xe1 and
  the residues which form the Xe1 pocket for the
  Xe1$\leftrightarrow$Xe2 transition. The average structure of Xe1 for
  the first umbrella of the forward direction, and for the last
  umbrella of the backward direction is compared. Both of these
  averaged structures correspond to $r_{\rm c}$ = 0, for their
  respective directions. The C$_{\alpha}$ of the residues forming Xe1
  pocket are superimposed. The RMSD of the heavy atoms of Xe1 pocket
  is presented.}
\begin{tabular}{|c|c|}
\hline
Residue & RMSD \\
\hline
Leu89 &  1.25 \\
His93 & 0.71 \\
Leu104 & 1.74 \\
Phe138 & 1.90 \\
Ile142 & 1.21 \\
Tyr146 & 0.68 \\
Xe1 & 1.33 \\
\hline
\end{tabular}
\flushleft
    
    \label{sitab2}
\end{table}

%\bibliography{n3p}